\documentclass[]{aa} % twocolumn option is crucial
\usepackage{txfonts}
\usepackage{natbib}
\usepackage{graphicx, xspace}

\usepackage[switch]{lineno}

\makeatletter

\newcommand{\Rmnum}[1]{\expandafter\@slowromancap\romannumeral #1@}
\makeatother

\usepackage[dvips]{color}

\newcommand{\new}[1]{\textbf{\blu{#1}}}

\renewcommand{\new}[1]{{#1}}					% no bold face

\def\gpsc{$\gamma$\,Psc\xspace}
\def\ttau{$\theta^1$\,Tau\xspace}
\def\deri{$\delta$\,Eri\xspace}
\def\Kepler{\textit{Kepler}\xspace}

\def\Hermes{\textsc{Hermes}\xspace}
\def\hermes{\textsc{Hermes}\xspace}
\def\Hides{\textsc{Hides}\xspace}
\def\hides{\textsc{Hides}\xspace}
\def\Coralie{\textsc{Coralie}\xspace}
\def\coralie{\textsc{Coralie}\xspace}
\def\Sophie{\textsc{Sophie}\xspace}
\def\sophie{\textsc{Sophie}\xspace}
\def\Sarg{\textsc{Sarg}\xspace}

\def\num{$\nu_\mathrm{max}$\xspace}
\def\dnu{$\Delta\nu$\xspace}
\def\dn1{$\delta\nu_{01}$\xspace}
\def\dn2{$\delta\nu_{02}$\xspace}
\def\sun{\hbox{$_\odot$}\xspace}

\begin{document}

%\linenumbers
%\modulolinenumbers[5]

\title{Detection of solar-like oscillations in the bright red giant stars \gpsc and  \ttau from a 190-day high-precision spectroscopic multisite campaign\thanks{Based on observations made with the \hermes spectrograph mounted on the 1.2\,m Mercator Telescope at the Spanish Observatorio del Roque de los Muchachos of the Instituto de Astrof'sica de Canarias; the CORALIE spectrograph mounted on the 1.2\,m Swiss telescope at La Silla Observatory, the HIDES spectrograph, mounted on the 1.9\,m telescope at Okayama Astrophysical Observatory, NAOJ, the MOST space telescope, and 
and observations made with ESO Telescopes at the La Silla Paranal Observatory under program ID 086.D-0101.}}
\authorrunning{P.\,G. Beck et al.}
\titlerunning{Detection of solar-like oscillations in the bright red giant stars \gpsc and  \ttau}

\author{P.\,G.~Beck\inst{1},  
E.~Kambe\inst{2}, 
M.~Hillen\inst{1}, 
E.~Corsaro\inst{1}, 
H.~Van\,Winckel\inst{1},
E.~Moravveji\inst{1}, 
J.~De Ridder\inst{1}, 
S.~Bloemen\inst{1,3},  
S.~Saesen\inst{4}, 
P.~Mathias\inst{5}, 
P.~Degroote\inst{1}, 
T.~Kallinger\inst{6},  
T.~Verhoelst\inst{7}, 
H.~Ando\inst{8}, 
F.~Carrier\inst{1}, 
B.~Acke\inst{1}, 
R.~Oreiro\inst{9}, 
A.~Miglio\inst{10}, 
P.~Eggenberger\inst{4}, 
B.~Sato\inst{11}, 
K.~Zwintz\inst{1}, 
P.\,I.~P\'{a}pics\inst{1}, 
P.~Marcos-Arenal\inst{1}, 
S.\,A.~Sans\,Fuentes\inst{1}, 
\\V.\,S.~Schmid\inst{1}, 
C.~Waelkens\inst{1}, 
R.~\O stensen\inst{1}, 
J.\,M.~Matthews\inst{12},
M.~Yoshida\inst{13}, 
H.~Izumiura\inst{2}, 
H.~Koyano\inst{2}, 
\\S.~Nagayama\inst{8}, 
Y.~Shimizu\inst{2}, 
N.~Okada\inst{8}, 
K.~Okita\inst{2}, 
A.~Sakamoto\inst{2}, 
T.~Yamamuro\inst{14},
C.~Aerts\inst{1,3}}

%\offprints{Paul G. Beck}
%\mail{paul.beck@ster.kuleuven.be}
\date{submitted 11 November 2013; accepted 16 July 2014}

\institute{Instituut voor Sterrenkunde, KU Leuven, 3001 Leuven, Belgium %1
\email{paul.beck@ster.kuleuven.be}
\and Okayama Astrophysical Observatory, National Astron. Obs. of Japan, Kamogata, Asakuchi, Okayama 719-0232, Japan %2
\and Department of Astrophysics, IMAPP, Radboud University Nijmegen, PO Box 9010, 6500 GL Nijmegen, The Netherlands %2
\and Observatoire de Gen\`eve, Universit\'e de Gen\`eve, Chemin des Maillettes 51, 1290 Sauverny, Switzerland %4
\and Universit\'e de Toulouse; UPS-OMP; IRAP; /
CNRS; IRAP; 57, Avenue d'Azereix, BP 826, F-65008 Tarbes, France%5
\and Institut f\"ur Astronomie der Universit\"at Wien, T\"urkenschanzstr. 17, 1180 Wien, Austria %6
\and Synergistic Exploitation of Atmospheric Data, Belgian Institute for Space Aeronomy, Brussels, Belgium%7
\and National Astronomical Observatory of Japan, 2-21-1 Osawa, Mitaka, Tokyo 181-8588, Japan %8
\and Instituto de Astrof'sica de Andaluc\'ia, Glorieta de la Astronom\'ia s/n, 18009, Granada, Spain%9
\and School of Physics and Astronomy, University of Birmingham, UK %10
\and Department of Earth and Planetary Sciences, Tokyo Institute of Technology, 2-12-1 Ookayama, Meguro, Tokyo 152-8551, Japan %11
\and  Department of Physics \& Astronomy, University of  British Columbia, Vancouver, V6T 1Z1, Canada%12
\and Hiroshima Astrophysical Science Center, Hiroshima University, 1-3-1 Kagamiyama, Higashi-Hiroshima 739-8526, Japan %13
\and OptCraft, 3-6-18 Higashi-Hashimoto, Midori-ku, Sagamihara 252-0144, Japan %14
}

\abstract
%\textbf{Context: }
{Red giants are evolved stars which exhibit solar-like oscillations.
 Although a multitude of stars \new{have been} observed with space telescopes, only a handful of  red-giant stars were targets of spectroscopic asteroseismic observing projects.}
%\textbf{Aims: }
{We search for solar-like oscillations in the two bright red-giant stars \gpsc and \ttau from time series of  ground-based spectroscopy and determine the frequency of the excess of oscillation power \num and the mean large frequency separation \dnu for both stars. Seismic constraints on the stellar mass and radius will provide robust input for stellar modelling.}
%\textbf{Methods: }
{The radial velocities of \gpsc and \ttau were monitored for 120 and 190 days, respectively.
Nearly 9000 spectra were obtained.
To reach the accurate radial velocities, we used simultaneous thorium-argon and iodine-cell calibration of our optical spectra. In addition to the spectroscopy, we acquired VLTI observations of \gpsc for an independent estimate of the radius. Also 22 days of observations of \ttau with the MOST-satellite were analysed.}
%\textbf{Results: }
{The frequency analysis of the radial velocity data of \gpsc revealed an excess of oscillation power around 32\,$\mu$Hz and a large frequency separation of 
4.1$\pm$0.1\,$\mu$Hz. \ttau exhibits oscillation power around 90\,$\mu$Hz,  
with a large frequency separation of 6.9$\pm$0.2\,$\mu$Hz. 
Scaling relations indicate that \gpsc is a star of about 1\,M\sun and 10\,R\sun. \ttau appears to be a massive star of about 2.7\,M\sun and 10\,R\sun. 
The radial velocities of both stars were found to be modulated on time scales \new{much longer than the oscillation periods}.}
%\textbf{Conclusions: }
{The estimated radii from seismology are in agreement with interferometric observations and also with estimates based on photometric data. While the mass of \ttau is in agreement with results from dynamical parallaxes, we find a lower mass for \gpsc than what is given in the literature. The long periodic variability agrees with 
the expected time scales of rotational modulation.}

\keywords{Asteroseismology, Stars: red giant, Stars: rotation, techniques: spectroscopic, techniques: photometric, techniques: interferometric; Stars: individual: $\gamma$\,Psc (HD\, 219615), $\theta^1$\,Tau (HD\,28307), $\delta$\,Eri (HD\, 23249),  open clusters and associations: individual: Hyades}

\maketitle

\section{Introduction}
\begin{table*}
\tabcolsep=5pt
\centering
\caption{\label{tab:literatureValues}Fundamental parameters and computed asteroseismic characteristics of the campaign targets. \label{tab:fundpara}}
\begin{tabular}{llllllllllllllll}
\hline\hline
Star& Sp. & RA& DE&V&$\pi$&$v\sin i$&T$_{\rm eff}$&[Fe/H]& Lum & Radius & Mass & $\nu_{\rm max}$ &$\Delta\nu$  &P$_{\rm rot}$\\
Name  &   Typ  & [h~m] & [$^\circ$ ']
    &	[mag]	&	[mas]	&	[km\,s$^{-1}$] & 	[K] &[dex] &	[L$_\odot$] & [R$_\odot$] & [M$_\odot$] & [$\mu$Hz]   & [$\mu$Hz] & [days]\\
\noalign{\smallskip}
\hline
\noalign{\smallskip}
$\gamma$~Psc & G9III &  23 17 & +03 17 & 3.7 & 23.6$\pm$0.2&4.7 & 4940&-0.62 & 63 &10.4  & 1.87 &  60 &  5.7   & 112.6\\
$\theta^1$\,Tau & K0III & 04 29 & +15 58 & 3.8 & 21.1$\pm$0.3 &4.2& 5000&+0.10 & 69 & 11.4 & 2.8$\pm$0.5 & 76 &  6.1   & 138.2\\  
\hline
\end{tabular}

\tablefoot{The coordinates (RA, DE) are given for the epoch 2000.0. The spectral type (Sp), luminosity class were adopted from the \textsc{Hipparcos} catalogue; parallaxes from \cite{vanLeeuwen2007}. 
The effective temperature (T$_{\rm eff}$), projected rotational surface velocity ($v\sin i$) was taken from \cite{hekker2007},  the metallicity [Fe/H] from \cite{Takeda2008}.
The luminosity (Lum) and stellar radius were obtained from the photometric calibration following \cite{Flower1996}. The mass for \gpsc and \ttau were taken from \cite{Luck2007} and \cite{Lebreton2001}, respectively.  The estimated frequency of the maximum oscillation power excess, \num, the large frequency separation $\Delta\nu$ and surface rotation period P$_{\rm rot}$ were calculated, based on the radius from the photometric calibration and the mass estimates from the literature. Uncertainties for all values  from literature are given when provided.}\bigskip

\tabcolsep=6pt
\caption{\label{tab:obsLog}\label{tab:observatories} 
Observatories participating in the campaign. }
\begin{tabular}{cccccccc|c|l}
\hline\hline
Instrument & Technique &Telescope &Location  & Calibr. & R & N$_{\gamma\,Psc}$ & N$_{\theta^1\,Tau}$ & Spectra & Observer\\ \hline
\textsc{Hermes} & Spec. & 1.2\,m Mercator & La Palma/Spain &   ThArNe & 46000 	& 1339	& 1329 	& 2668 & PGB, MH, SB, et\,al. \\
\textsc{Coralie}& Spec. &1.2\,m Euler & La Silla/Chile &   ThAr & 62000 	& 1949	& 792	& 2741 & PGB, SS\\
\textsc{Hides} & Spec. &1.88\,m telesc. & Okayama/Japan & Iodine & 50000	& 1306	& 1284	& 2590 & EK, AH, PGB\\
\textsc{Sophie} & Spec. &1.93\,m telesc. & OHP/France & ThAr 	& 75000 & 485	& 360	& 845 & PM\\
\textsc{Sarg}& Spec. &3.58\,m TNG & La Palma/Spain &  Iodine & 46000 &-&-&-&RO\\ \hline

\textsc{Amber} & Interf. & VLTI&Paranal/Chile &  - & 30 &-&-&-&SB\\
\hline
\end{tabular}

\tablefoot{The first four columns give the instrument, technique, telescope and its observing site used for the observations.
The abbreviations ThAr, ThArNe and Iodine of the applied calibration methods stand for simultaneous ThAr or ThArNe reference and Iodine 
cell, respectively. The spectral resolution $R$ of the utilised observing mode is given. The next three columns list 
the numbers of spectra taken for \gpsc, \ttau and in total per telescope. 
The last column gives the initials of the main observers at the telescope. All observers are listed as coauthors.}
\end{table*}

Solar-like oscillations are excited stochastically by motions in the convective envelope of stars. For low-mass stars, they are found in all evolutionary states, between the main sequence and horizontal branch of helium-core burning stars \citep[e.g.][]{Leighton1962, frandsen2002, Carrier2003, Hekker2009, Chaplin2011, huber2011, kallinger2012, Mosser2013b} and were even detected in the M5 super giant $\alpha$\,Her \citep{Moravveji2012}. These very characteristic oscillations lead to a nearly regular spaced comb-like pattern in the power spectrum. 
It was shown by \cite{Deubner1975} that this ridge structure is governed by the degree of oscillation modes and resembles the predictions made by \cite{Ando1975}. %The theoretically described by \cite{tassoul1980}. 
Empirically, the frequency patterns of solar-like oscillations were described through scaling relations by \citet{kjeldsen1995}. Since then, these relations have been tested and revised from large sample studies based on high-precision space photometry of red giants in clusters and in eclipsing binaries besides single stars \citep[e.g.][respectively]{corsaro2012,frandsen2013, kallinger2010}. 
Indications of non-radial oscillation modes were found in the variations of the absorption lines of bright red giants \citep{hekker2006, hekkerAerts2010} but firmly established in a large set of red giants observed with the \textsc{CoRoT} 
satellite \citep{deRidder2009}. The identification of dipole mixed modes  extended the sensitivity of the seismic analyses also towards the 
core  of evolved stars \citep{beck2011,bedding2011,mosser2011a}. 
The analysis of solar-like oscillations enabled us to unravel many open questions on stellar structure and evolution, such as constraining 
the internal rotational gradient \citep{elsworth1995,beck2012,deheuvels2012} or determining the evolutionary status in terms of nuclear 
burning of a given red-giant star \citep{bedding2011,mosser2011a}.

Spectroscopic time series offer a different look at stars, as spectroscopy and photometry have different mode sensitivities because modes with an odd spherical degree $l$ exhibit a better visibility in spectroscopy than in photometry \citep{Aerts2010}. Radial velocity data is less effected by granulation noise, providing a wider frequency range that can be investigated \citep{Aerts2010}.
In addition, most targets of space observations are faint. As typical target stars for spectroscopy are bright, they are well studied and many complimentary parameters are found in the literature. Furthermore, experience has shown that the long term stability is a major issue for space missions 
and it is very challenging to constrain weak signals on the order of several tens of days \citep[e.g.][]{beck2014a}. With robust calibration 
techniques, we can also study variations on the order of more than 100\,days and longer from stable spectrographs.

\begin{figure*}[t!]
\centering
\includegraphics[width=\textwidth]{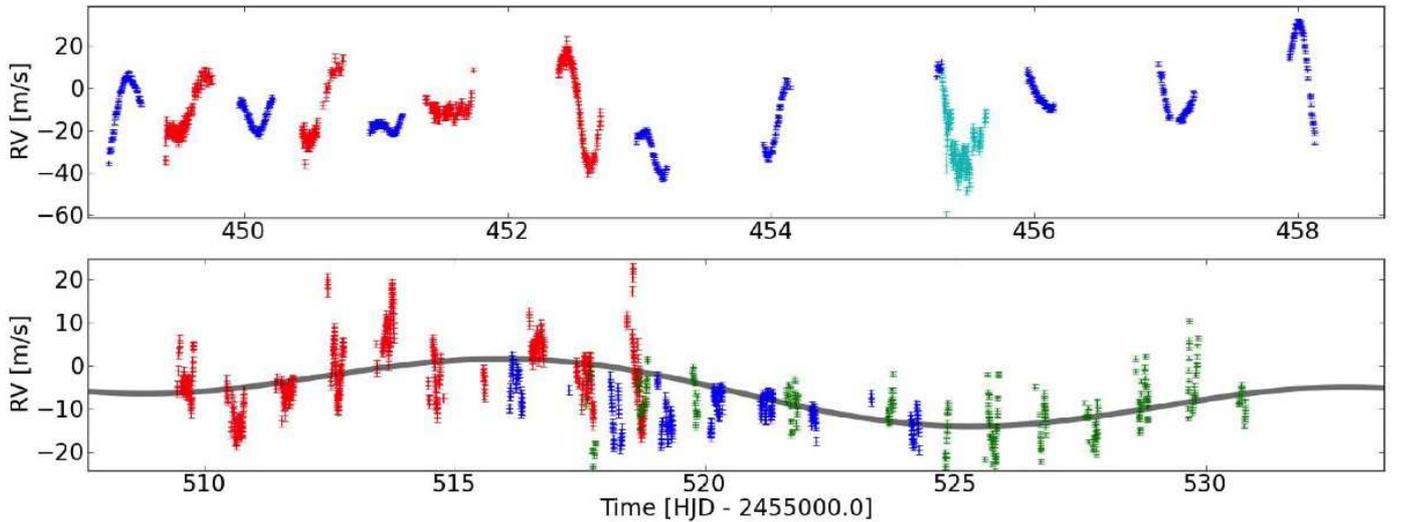}
\caption{\label{fig:rvShorteTimescales}
\new{Example} of radial velocity measurements of \gpsc in September 2010 (top panel) and \ttau in November 2010 (bottom panel). The data sets of \hermes, \hides, \coralie and \sophie are shown in \new{red,} blue, \new{green,} and cyan, respectively. \new{The full data set is depicted in Fig.\ref{fig:curvesRV}. The radial velocities of \ttau have been detrended for the binary orbit, shown in Fig.\,\ref{fig:binaryTTAU}. \new{The gray line indicates the 16.57\,d modulation of the radial velocities, found in \ttau.}}
%The time series of \gpsc in the top panel show a clear variation with a characteristic time around 8.7h ($\sim$32\,$\mu$Hz) present in all data sets. The solid line in the bottom panel shows the long-term variation, described by the two frequencies in Table 7, which is also consistently present in all data sets.
}\end{figure*}

Although space-photometry is available for stars in all phases on the red-giant branch, no spectroscopic time series have deliberately been 
obtained for stars in the red clump.  Most campaigns concentrate on stars close to or on the main sequence, as the higher oscillation frequencies 
do not require extensive time series to resolve the variation.  In this paper we describe the definition, data acquisition, and analysis of an 
extensive observational campaign, focused on the two red giants \gpsc and \ttau, from high-precision spectroscopy. We also compare our results 
to interferometrically determined radii. 

% ======================================================
\section{Project strategy and observations}

\begin{figure*}[t!]
\includegraphics[width=\textwidth%,height=110mm
]{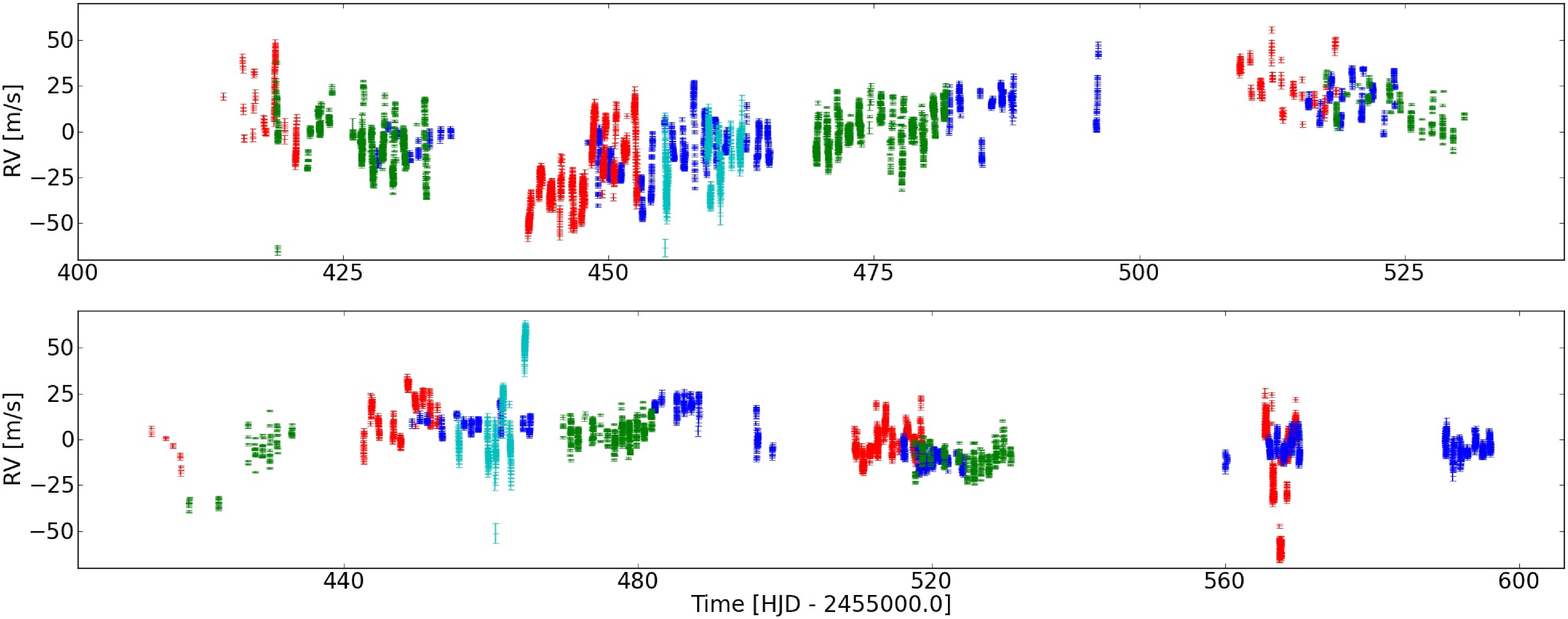}
\caption{\label{fig:curvesRV}
Radial velocity curve of \gpsc (top panel) and \ttau (bottom panel). The radial velocities of \ttau have been detrended for the binary orbit, shown in Fig.\,\ref{fig:binaryTTAU}.  The red, blue, green, and cyan data points have been acquired with the \Hermes, \Hides, \Coralie, and \Sophie spectrograph, respectively.} 
%\end{figure*}
%\begin{figure*}[t!]
\includegraphics[width=\textwidth]{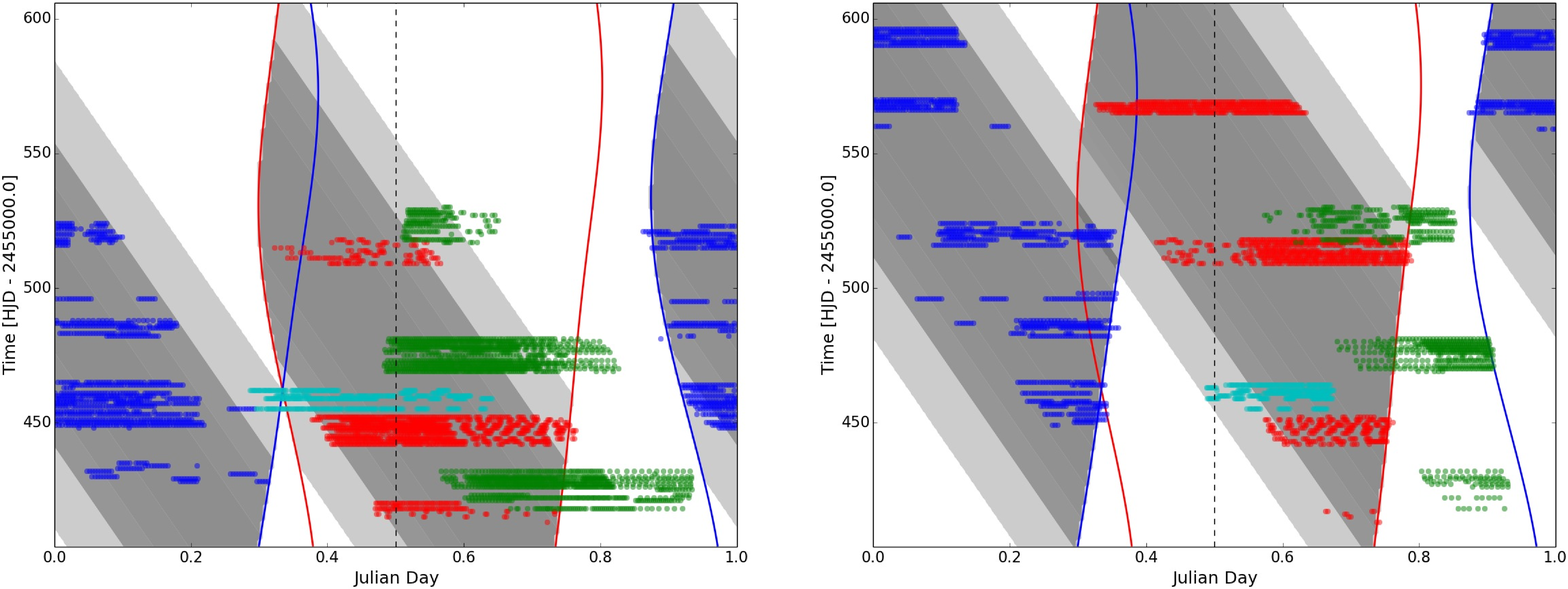}
\caption{\label{fig:observersEchelle}
\new{Distribution of the radial velocity measurements of \gpsc (left panel) and \ttau (right panel).  The same color code as in Fig.\,\ref{fig:rvShorteTimescales}.
For \Hides and \Hermes, also the value of the airmass X at which the target was observed is shown. Light gray shaded areas indicate 2.5\,$<$\,X$<$\,3 and dark gray X\,$<$2.5. 
The solid blue \new{and red} lines indicate the nautical dawn for \hides and \hermes, respectively.}
%The \new{vertical} dashed line corresponds to 0 hrs universal time (UT). 
}\end{figure*}

Ground-based time series severely suffer from the contamination of alias-frequencies. However, it has been shown in numerous photometric and spectroscopic campaigns that the coordinated, time-resolved observation of an object with telescopes at several, well distinguished geographic longitudes is highly improving the structure of the spectral window which originates from the gaps \citep[e.g.,][for various types of pulsators]{Winget1990, Handler2004, Breger2006, arentoft2008, Desmet2009, Kolenberg2009, Saesen2010}. \new{To further improve the structure of the power spectrum, several recent works have modelled the effects of an incomplete duty cycle on solar-like oscillations for ground-based, multi-site observations and also for space observations with the \Kepler space telescope  \citep[][respectively]{Arentoft2014,Garcia2014b}. However, our data consists of individual datasets, originating from different telescopes and different calibration techniques with large gaps \new{(Fig.\,\ref{fig:rvShorteTimescales} and \ref{fig:curvesRV}). The structure of the dataset is visualized in the graphical observing log in Figure \ref{fig:observersEchelle}.} We therefore decided not apply such techniques, in order to avoid smearing or correlating systematic effects between datasets.} 

\subsection{Target selection \label{sec:targetSelection}}
We organised an extensive observational campaign, focusing on two red-giant stars using spectrographs well distributed over geographical 
longitude and capable of meter-per-second precision. As such instruments are not numerous while located on both hemispheres, target stars 
had to be close to the celestial equator. To achieve high-precision radial velocity measurements with a high temporal resolution on the 
order of a few minutes, we limited our search to giants brighter than 4$^{th}$ magnitude in the visual.

Starting from the \textsc{Hipparcos}-catalogue, all stars, which did not fit the limitations in magnitude, colour index or declination were 
excluded. As a goal of this campaign is to study the effects of rotation on oscillation in red giants \citep{beck2010}, stars with an unknown 
projected rotational velocity or with an averaged projected rotational velocity lower than 3 km\,s$^{-1}$ were excluded from the target selection. 
For the remaining candidates a photometric calibration, following the grid of \cite{Flower1996} has been carried out, obtaining the bolometric correction, effective temperature, and radius for each star. The \textsc{Hipparcos}-parallaxes from \cite{vanLeeuwen2007} 
have been used for the distance determination. To check if the frequency spectra would be suitable for this project the asteroseismic parameters of solar-like oscillations of 
the remaining targets were computed following \cite{kjeldsen1995}. 

To use the observing time efficiently, we were searching for a target combination, that encompasses two targets, separated by approximately 5 hours 
in right ascension. After testing several target combinations, we finally selected the {G9III} star $\gamma$\,Psc (HD\,219615) and the K0III star $\theta^1$\,Tau 
(HD\,28307) as targets for our campaign.

\subsection{Characteristics of \gpsc and \ttau}

Both targets are evolved stars which are expected to exhibit solar-like oscillations. An overview of the fundamental parameters of these stars from the literature is given in Table\,\ref{tab:literatureValues}. \gpsc is a single field star 
which was studied frequently in the past using photometric and spectroscopic measurements \citep[e.g.][]{Luck2007,hekker2007}. The most important feature of \gpsc with respect to the present study is its subsolar iron abundance (Table.\,\ref{tab:literatureValues}).

\ttau is one of the four giants in the Hyades cluster and a well studied binary \citep[e.g.][]{Lunt1919, Torres1997, Lebreton2001, Griffin2012}. The Hyades have a slightly higher metallicity than the Sun with Fe/H = +0.1 \citep{Takeda2008}. By comparing theoretical isochrones matching the measured Helium and metal content of the cluster, \cite{Perryman1998} found a main-sequence turn off point of $\sim$2.3\,M$_\odot$ which corresponds to a cluster age of 625$\pm$50 Myr. Therefore only the most massive stars are already in more advanced evolutionary stages after the main sequence. \new{Very recently, \cite{Auriere2014} reported on the detection of a magnetic field in \ttau whose unsigned longitudinal component peaked to 3$\pm$0.5\,G. From their measurements they find indications for a complex surface magnetic field structure.}

Interferometric observations allow for a radius determination, independent of seismology. Interferometric observations of the Hyades' giants were published by \cite{Boyajian2009}, reporting a radius of 11.7$\pm$0.2\,R$_\odot$ for \ttau. 
\new{For \gpsc no such observations \new{are} reported in the literature. We therefore obtained interferometric observations of this star with the \textsc{Amber} instrument at the Very Large Telescope Interferometer (VLTI, cf. Section\,\ref{sec:interferometry}).}

\subsection{Participating observatories}
The main part of the campaign took place from August 2010 through January 2011 and utilised 5 different telescopes.We monitored \gpsc and \ttau 
for $\sim$120\,d and $\sim$190\,d, respectively, during which in total 8844 spectra were taken with the  \Hermes, \Coralie, \Sophie,  \Hides and \Sarg spectrographs. 

The majority of data was obtained with the \textsc{Hermes} spectrograph \citep{raskin2011} mounted at the 1.2\,m \textsc{Mercator} telescope 
on La Palma, Canary Islands, the \textsc{Coralie} spectrograph \citep[e.g.][]{Queloz2000} operated from \textsc{Mercator}'s twin 1.2\,m \textsc{Euler} 
telescope on La Silla, Chile and the \textsc{Hides} spectrograph \citep{Izumiura1999} at the Okayama 1.88\,m telescope, Japan. We used the commissioning time of the fibre-fed high-efficiency \new{observing} mode of \hides for the campaign \citep{Kambe2013}. Additional data was acquired with the \textsc{Sophie} spectrograph \citep{Bouchy2009} at the 1.93\,m telescope at the Observatoire de Haute-Provence (OHP) in the south of France. 
During the phase of best visibility for both targets, we obtained spectroscopic data with the \textsc{Sarg} spectrograph \citep{SARG2001} mounted at the 3.58\,m Telescopio Nazionale Galileo (TNG), on La Palma. Due to problems in the reduction of the \Sarg data, they could not be included in the present analysis. \new{An overview of the applied calibration techniques, the observers and the number of spectra taken with each instrument is given in Table\,\ref{tab:obsLog}.}

Additionally, for \gpsc we applied for and obtained two half nights with the \textsc{Amber} instrument at the Very Large Telescope Interferometer (VLTI), using the 1.8m auxiliary telescopes (See Section\,\ref{sec:interferometry}). 

In 2007 the MOST satellite \citep{Walker2003} was used to search for photometric variability in the Hyades' giants. The MOST team kindly provided us with the previously unpublished light curve of \ttau, covering 22 days to compare the photometric and the radial velocity variations (See Section\,\ref{sec:MOST}).

% ======================================================
\subsection{Radial velocity measurements and data set merging}

In order to achieve ms$^{-1}$ precision in our radial velocity measurements, we utilised the available simultaneous calibration techniques. The \textsc{Hermes}, \textsc{Coralie} and \textsc{Sophie} spectrographs use Thorium-Argon (ThAr) or Thorium-Argon-Neon (ThArNe) reference lamps, from which light is simultaneously fed to the spectrograph through a reference fibre. On the CCD, the stellar and the reference spectrum are alternating in position. Radial Velocity information was derived from the spectra using the instrument pipelines. \textsc{Hides}  use the absorption spectrum of an Iodine (I$_2$) cell, superimposed on the stellar spectrum in a wavelength range of 500 to 620\,nm. \hides data was reduced with IRAF and the radial velocity was estimated with the method described in \cite{Kambe2008} by a member of the \Hides team.

\begin{figure}[t!]
\includegraphics[width=0.5\textwidth]{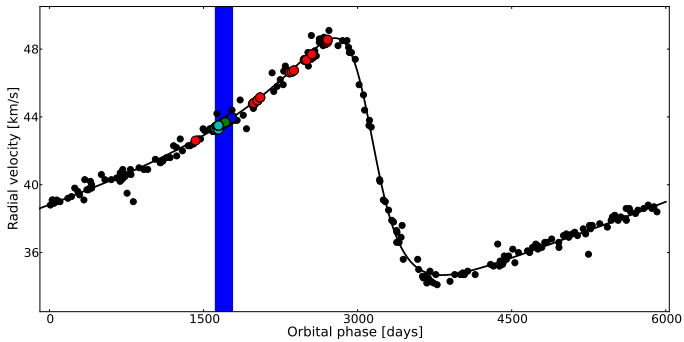}
\caption{\label{fig:binaryTTAU}
Orbital velocity phase diagram of \ttau. Data by \cite{Griffin2012} between the years 1969 and 2010 is shown as black dots. The duration of our multisite asteroseismology campaign is marked with the blue bar. Red dots mark the values from the ongoing radial velocity monitoring with the \Hermes spectrograph. The values are reported in Table\,\ref{tab:ttauHermesRV}.} 
\end{figure}

The mean radial velocity of each data set was subtracted from the observations of \gpsc to merge the results from all 4 different telescopes, in order to obtain a consistent radial velocity curve. As \ttau is a binary, we subtracted an orbital model (Table\,\ref{tab:orbitalParametersTTAU}, \new{Figure\,\ref{fig:binaryTTAU}}), derived from the data set of \cite{Griffin2012}, equally distributed over 40 years (from 1969 through 2010). As we have no overlap of data, except for about 1 hour between \Hermes and \Coralie and only a few radial velocity standards were observed, we manually corrected the offsets between the radial velocities from different spectrographs. 

\new{In addition to the data of Griffin, we also monitored the radial velocity of \ttau with \hermes between early 2010 up to late 2013. The radial velocity values from \cite{Griffin2012} as well as from our monitoring, listed in Table\,\ref{tab:ttauHermesRV} are compared in the phase diagram, depicted in Figure\,\ref{fig:binaryTTAU}.}

\begin{table}[t!]
\tabcolsep=2.3pt
\centering
\caption{\label{tab:orbitalParametersTTAU} 
Orbital parameters for \ttau. }
\begin{tabular}{cccccc}\hline\hline
P  & T$_{0}$ & $e$ & $\omega$ & K & $\gamma$ \\
$[$d$]$ & [d] &  & [rad] & [km\,s$^{-1}$] & [km\,s$^{-1}$] \\ \hline
5927$\pm$7 & 2439132$\pm$12 & 0.56$\pm$0.01 & 1.20$\pm$0.01 & 6.99$\pm$0.05 & 40.23$\pm$0.03 \\ \hline
\end{tabular}

\tablefoot{The columns give the orbital period $P_{\rm orbit}$, the zero point T$_0$, the eccentricity $e$, the argument of periastron $\omega$, the radial velocity 
amplitude K, and the systemic velocity $\gamma$.} \\ \bigskip
%\end{table}
%\begin{table}[t!]
\tabcolsep=5pt
\caption{\label{tab:ttauHermesRV}
Radial velocities of \ttau from \textsc{Hermes} observations in the high-resolution observing mode.}
\begin{center}
\begin{tabular}{ccc}
\hline\hline
HJD & RV & error$_{\rm RV}$\\
$[$d$]$ & [km\,s$^{-1}$] & [km\,s$^{-1}$] \\ \hline
2455241.38046	&	41.645	&	0.001\\ 
2455799.74878	&	43.793	&	0.002\\ 
2455806.74723	&	43.826	&	0.001\\
2455809.75936	&	43.865	&	0.001\\ 
2455842.75045	& 	44.008	&	0.001\\
2455870.65140	& 	44.186	&	0.001\\
2456157.69844	&	45.671	&	0.001\\
2456179.69122	&	45.704	& 	0.001\\
2456305.41881	&	46.401	&	0.001\\
2456317.38334	& 	46.396	&	0.001\\
2456322.38404	&	46.398	&	0.001\\ 
2456374.35806	&	46.738	&	0.002\\
2456507.72800 	& 	47.417	& 	0.002\\
2456510.72944	&	47.419  	&	0.001\\
2456518.71171	&	47.494	&	0.002\\
2456523.74224	& 	47.481	&	0.002\\
2456526.74136	&	47.590	&	0.002\\
2456576.53391  &	47.681	&	0.001\\
\hline
\end{tabular} 
\end{center}
\tablefoot{The radial velocities and their formal uncertainties were determined through the cross-correlation pipeline for the \Hermes spectrograph. The average of consecutive measurements is given. The observations were obtained between February 2010 and October 2013.}
\end{table}%

\new{An inspection of the combined data subsets, as depicted in Figure\,\ref{fig:rvShorteTimescales}, shows that both stars exhibit radial velocity variations.
The colours of the data points show the contribution of the different observatories. Although from single-site data the long periodic variations of \gpsc could be seen as long periodic trends, it becomes evident from the combination of multi-site data that it is a clear oscillation signal. Figure\,\ref{fig:rvShorteTimescales} also suggests that \ttau oscillates at higher frequencies than \gpsc. The full radial velocity curves are shown in Fig.\,\ref{fig:curvesRV}.}

\new{Figure\,\ref{fig:observersEchelle} visualises the observational coverage of our campaign in an \'echelle diagram style. The vertical axis of both diagrams shows  the distribution of observations during the project from August 2010 to January 2011. The horizontal axis shows the time of the day as fraction of the Julian day. It therefore depicts the coverage of the multisite campaign. For a more intuitive understanding of the diagram, the airmass of the object and the laps of the nautical twilight are given for the geographical location of the \hides and \hermes spectrographs. }

For improving the agreement between different data sets while merging, and for increasing the signal-to-noise ratio in the amplitude spectrum, we normalised the uncertainties of the individual radial velocity measurements following the approach developed by \cite{Butler2004} and \cite{Bedding2007}, which was described and adopted by \cite{corsaro2012}.
The analysis consists of two steps. First, the individual uncertainties in the radial velocity measurements are rescaled in order to be consistent with the amplitude of the noise level in the amplitude spectra, $\sigma_\mathrm{amp}$, which is measured for each data set separately in a region far from the power excess. Second, a residual time-series is derived by a complete pre-whitening of the oscillation signal, in order to end up with only noise remaining in the data sets.

\begin{figure}[t]
\includegraphics[width=0.48\textwidth]{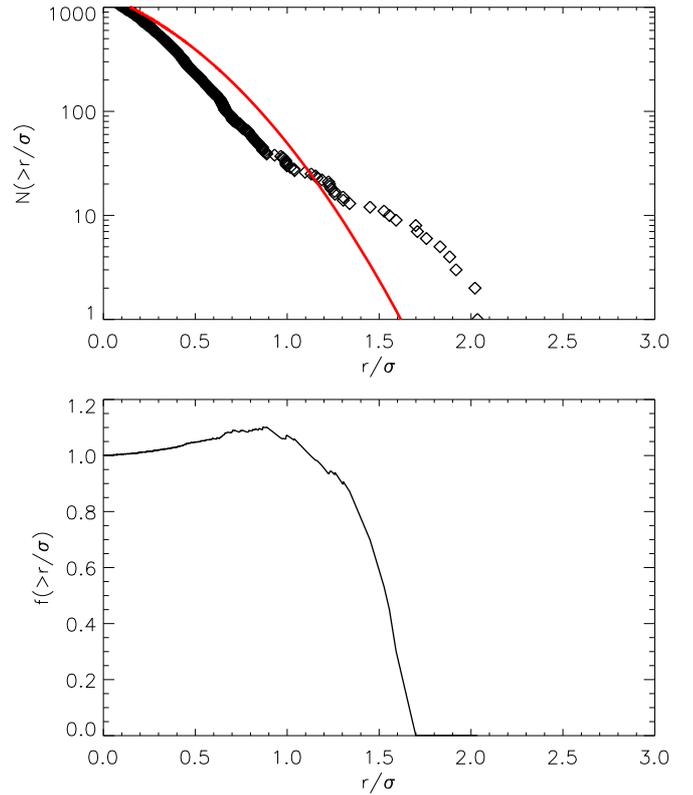}
\caption{\label{fig:cumulativeHistogramHermes} 
Cumulative histograms for \Hermes observations of \gpsc. 
Upper panel:  The diamonds represent the ratio $|r_i / \sigma_i |$ between the residuals in the pre-whitened time-series and the uncertainties of each radial velocity measurement.  The solid red curve depicts the result expected for Gaussian noise. 
The diagram in the lower panel shows the ratio between the observed and theoretical distribution of the uncertainties.}
\end{figure}

The results of this approach in the case of \Hermes data for \gpsc is depicted in Fig.\ref{fig:cumulativeHistogramHermes}. The upper panel shows the observed cumulative histogram of the ratio $|r_i / \sigma_i|$ of the residuals $r_i$ to the original uncertainty $\sigma_i$ in each radial velocity measurement (black diamonds) with the theoretical distribution expected for Gaussian noise (red line). The uncertainty of the individual radial velocity measurement is then adjusted by the ratio $f$ of the two distributions, which is shown in the lower panel. The ratio enhances the presence of a number of outliers, i.e. data points deviating from the expected distribution with $|r_i / \sigma_i| > 1.2$, which are down-weighted when computing the amplitude spectrum.

\begin{table}[t!]
\caption{\label{tab:accuracyPerMeasurement}
Accuracy per measurement of the radial velocity observations.}
\begin{tabular}{c|r|rr|rr}\hline\hline
& &\multicolumn{2}{c}{\bf original uncertainties}  & \multicolumn{2}{c}{\bf normalised uncert.} \\
{\bf \gpsc} & N & $\sigma_{\rm noise}$ &
$\sigma_{\rm meas.}$  & $\sigma_{\rm noise}$ & $\sigma_{\rm meas.}$\\
& &[m\,s$^{-1}$] & [m\,s$^{-1}$] & [m\,s$^{-1}$] & [m\,s$^{-1}$]\\

\hline
\Hermes	& 1316	& 0.3	& 7.3	& 0.3	& 7.4\\
\Coralie 	& 1946	& 0.2	& 7.2	& 0.2	& 6.6\\
\Hides 	& 1306	& 0.1	& 4.2	& 0.1	& 3.6\\
\Sophie	& 485	& 0.6	& 10.8	& 0.5	& 8.7\\					
\hline
All 		& 5053 &  0.1	& 7.0  	& 0.1	& 5.4 \\
\hline
\multicolumn{5}{c}{~}\\
\hline
{\bf \ttau} & N & $\sigma_{\rm noise}$ &
$\sigma_{\rm meas.}$  & $\sigma_{\rm noise}$ & $\sigma_{\rm meas.}$\\
& &[m\,s$^{-1}$] & [m\,s$^{-1}$] & [m\,s$^{-1}$] & [m\,s$^{-1}$]\\
\hline
\Hermes 	& 1286	& 0.1	& 3.5	& 0.1	& 4.1\\
\Coralie 	& 790	& 0.2	& 4.4	& 0.2	& 4.5\\
\Hides 	& 1283	& 0.1	& 2.3	& 0.1	& 2.1\\
\Sophie	& 360	& 0.6	& 9.7	& 0.4	& 6.1\\
\hline
All 	& 3719 &  0.1  	& 4.3	& 0.1 & 3.0 \\
\hline
\end{tabular}

\tablefoot{$N$ gives the number of data points finally used in the single-site data set. 
$\sigma_{\rm noise}$ reports the average noise amplitude in the frequency range between 
1000 and 1200\,$\mu$Hz.  $\sigma_{\rm meas.}$ is the average measurement uncertainty in a data set.}
\end{table}%

As a conclusion, we were able to reduce the noise level in the amplitude spectrum of the master data set from 
$\sigma_{\rm amp} = 0.09\,$m\,s$^{-1}$ to 0.06\,m\,s$^{-1}$ for $\theta^1$ Tau, and from $0.12$\,m\,s$^{-1}$ to $0.1$\,m\,s$^{-1}$ for $\gamma$ Psc.  
These values are the mean noise amplitude, derived for the frequency range 1000$-$1200\,$\mu$Hz of the power spectrum which is supposedly free from leakage of oscillation power. The values of the mean noise amplitude in this range and how they were modified through the normalisation of weights is shown in Table\,\ref{tab:accuracyPerMeasurement}. The frequency analyses of these data are presented in Sections\,\ref{sec:solarlikeOscillationsGpsc} and \ref{sec:solarlikeOscillationsTtau} and were performed on these adjusted residual radial velocity curves.

\subsection{Instrument commissioning of \textsc{Hermes} \label{sec:hermesComissioning}}

The campaign was also used for commissioning the fibre-fed high efficiency \new{observing} mode of the simultaneous ThArNe observing (low resolution with reference fibre, i.e. LRFWRF) of the \Hermes spectrographs.  In this \new{observing} mode, \Hermes is simultaneously fed from the telescope and a ThArNe calibration unit through 60\,$\mu$m fibres. The exposure times were chosen to achieve a signal-to-noise ratio of 150 in the blue part of the stellar spectrum. To avoid saturation of the emission line spectrum and an eventual contamination of the stellar spectrum through blooming of strong lines on the CCD chip, a neutral density filter regulates the intensity of the reference source. The obtained spectra were reduced with the instrument specific pipeline, extracting a 1D spectrum for the stellar and reference fibre. From each spectrum, the radial velocities of the star were derived through cross-correlation of the wavelength range between 478 and 653\,nm with a line-list template of the spectrum of Arcturus \citep{raskin2011}. The radial velocities are then corrected for the shift in radial velocity, determined from the cross correlation of the simultaneously exposed ThArNe spectrum with the reference ThArNe spectrum, obtained in the beginning of the night.

\subsubsection{\deri}
\begin{figure}[t!]
\centering
\includegraphics[width=0.55\textwidth]{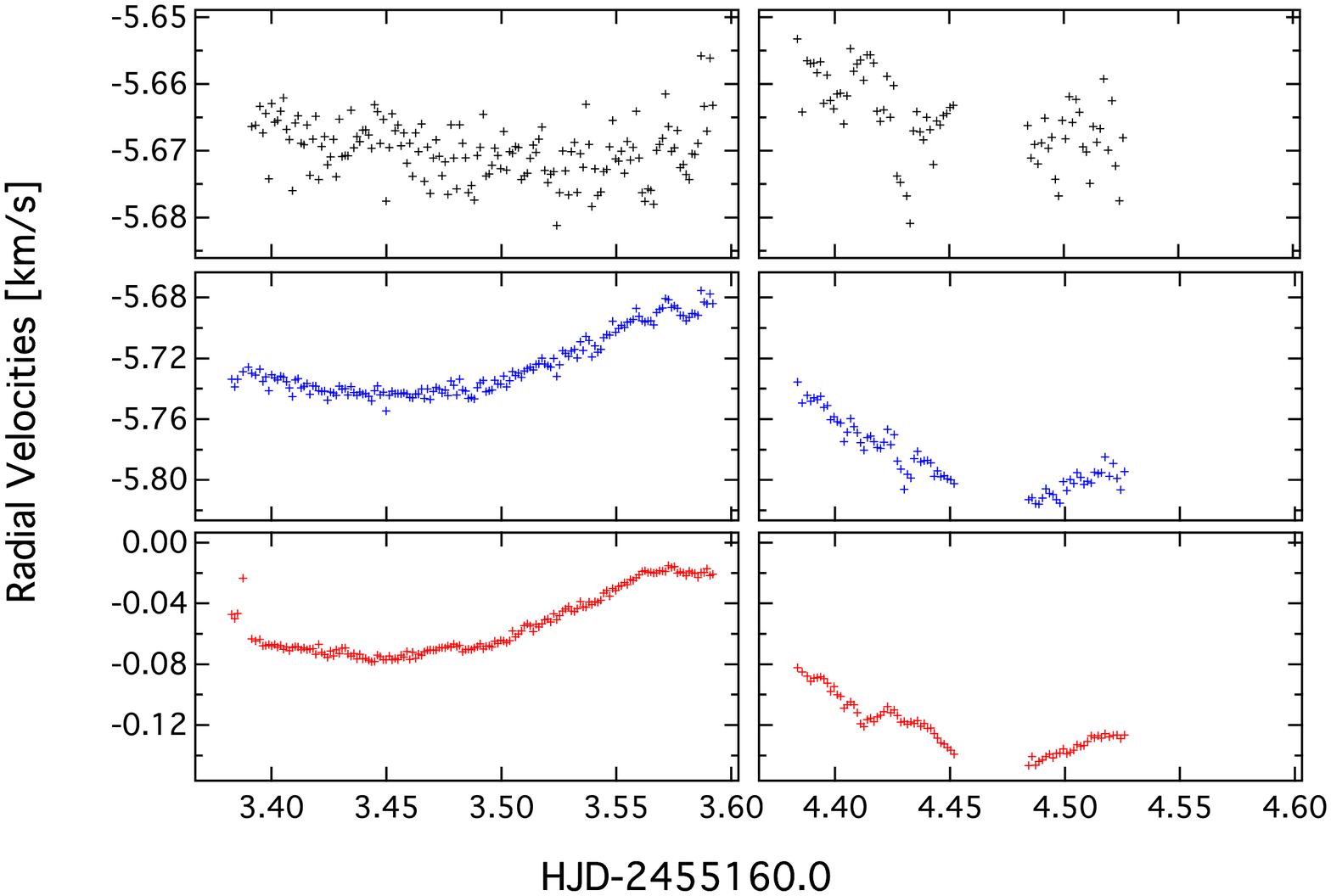}

\caption{\label{fig:DeltaEriRV}
Radial velocity of \deri. Top panel shows the radial velocity values corrected for radial velocity drifts. The middle panel shows the uncorrected velocity values from the stellar spectrum while the bottom panel gives the radial velocity drift of the simultaneous ThArNe with respect to the wavelength reference, obtained in the beginning of the night.}
\vspace{5mm}
\includegraphics[width=0.5\textwidth]{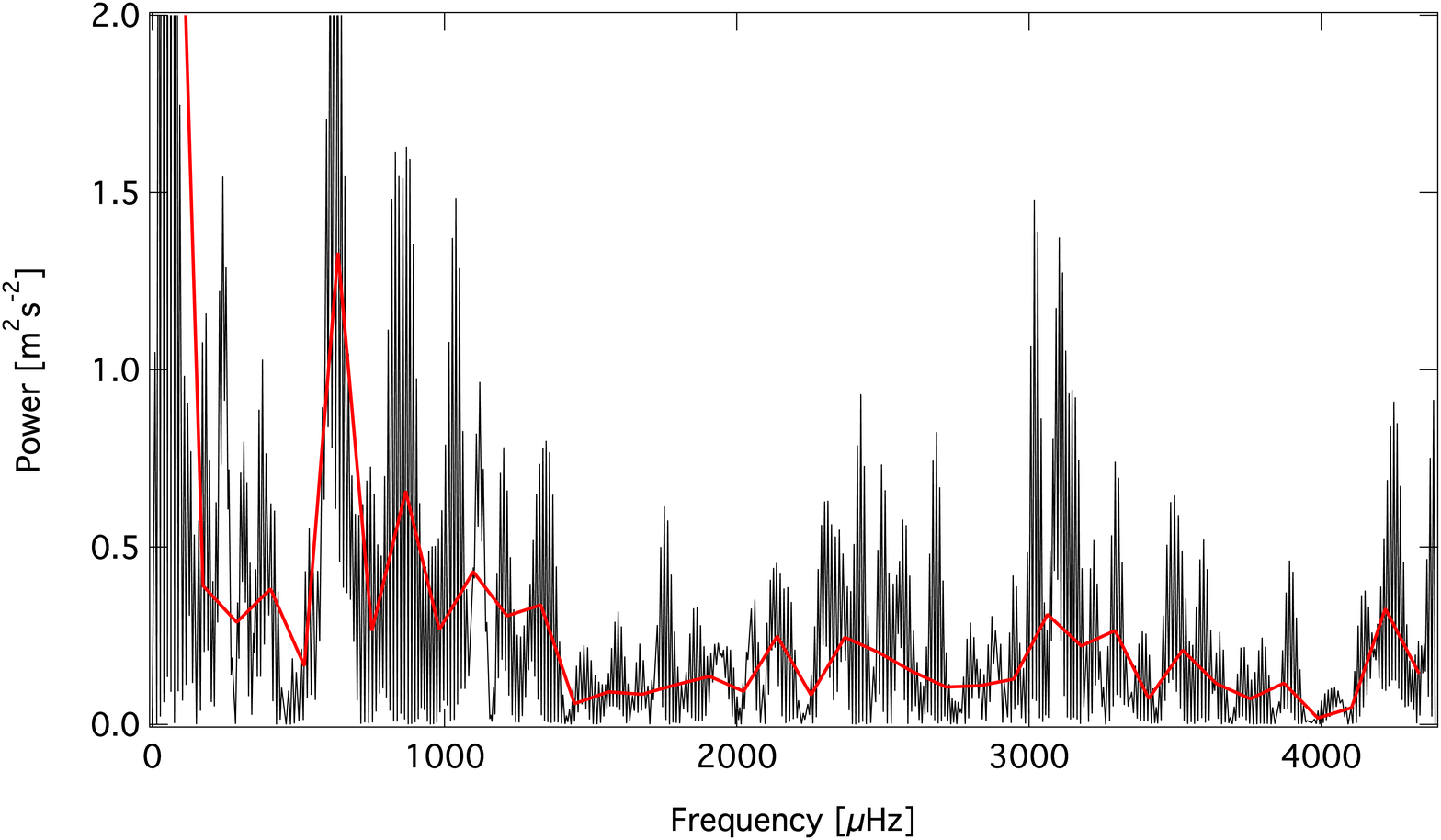}
\caption{\label{fig:DeltaEriPS}
Power spectrum for the RV measurements of \deri. The red line illustrates the running average noise level in the spectrum.}
\end{figure}

To test the performance and the accuracy of the  \new{observing} mode 
before the campaign, we have obtained observations of $\delta$\,Eri, a well known solar-like 
oscillator, exhibiting power between 500-900\,$\mu$Hz \citep{Carrier2003}. Our data set covers 14.4 hours in two consecutive nights in late November 2009. In total, 235 spectra have been obtained with an exposure time of 60 to 90 seconds. The resulting radial velocity time series and the corresponding power spectrum of \deri are shown in Figs.\,\ref{fig:DeltaEriRV} and \ref{fig:DeltaEriPS}.

From the comparison with earlier observations obtained by \cite{Carrier2003}, we find that the region between 1500 and 4000\,$\mu$Hz is free from contamination through oscillations and resembles the average amplitude $\sigma_{\rm noise}$, which originates from instrumental noise and other noise sources. Following the Parseval theorem, we obtain 
\begin{equation}
\label{eq:measurementAccuracy}
\sigma_{\rm measurement}=\sqrt{\frac{2 \cdot N}{\pi}}\cdot\sigma_{\rm noise},
\end{equation} the average noise amplitude of 0.24 m\,s$^{-1}$ in this region translates into an accuracy of about 2.9 m\,s$^{-1}$ per measurement which is more than sufficient to explore solar-like oscillations in red-giant stars. 
\new{These observations were obtained in an early phase of the commissioning.
As the instrument was improved since the observations of \deri,} \new{we did not find such noise in the data set of the campaign in 2010.} Therefore, the noise in this frequency region is likely be coloured noise, originating from instrumental effects. These observations of \deri thus provided us with a proof-of-concept for the main observing campaign.

\subsubsection{\gpsc and \ttau}
We performed the same analysis for \gpsc and \ttau~for the frequency range between 1000 and 1200\,$\mu$Hz on the campaign data, because we did not detect the signature of spectral leakage of oscillation power in this region. We calculated the noise amplitude for this region and estimated the accuracy per measurement from Eq.\,(\ref{eq:measurementAccuracy}) for individual data sets. An overview is given in Table\,\ref{tab:accuracyPerMeasurement}.

The numbers in Table\,\ref{tab:accuracyPerMeasurement} show that, although both stars were observed with the same instruments, the intrinsic scatter  in the power spectra of \gpsc is consistently 1.8 times higher than for \ttau. One possible explanation for this velocity jitter could be systematic effects as the accuracy of the measurements depend on how suited a star template for the cross correlation is for a given star. All pipelines used templates for K0 stars.
\cite{Carney2003} concluded that neither orbital motion nor pulsation can explain such velocity jitter and suggest star spots or activity as another possible explanation. 

% ======================================================

\section{Interferometric observations of \gpsc \label{sec:interferometry}}

The AMBER instrument \citep{Petrov2007} on the Very Large Telescope Interferometer (VLTI) was used to observe \gpsc in its low spectral resolution (LR) \new{observing} mode (R=30),
covering the J, H and K atmospheric bands. Observations were procured during the nights of 4 and 5 October 2010, using two triplets of Auxiliary Telescopes (ATs) 
that included stations A0 and K1, the longest baseline (128m) available at that time (see Table~\ref{table:obslogAMBER}). The VLTI first-generation 
fringe tracker FINITO \citep{leBouquin2008} was used to track the fringes.

\subsection{Reduction}

The data reduction was performed with the standard software \textit{amdlib} v3.0.5, 
provided by the Jean-Marie Mariotti Center (JMMC) \citep{2007AATatulli,2009AAChelli}. With \textit{amdlib}, first a P2VM (Pixel-To-Visibility-Matrix) is computed,
which is then used to translate the detector frames into interferometric observables. In our case, a typical exposure consisted of 1000 frames, with integration times
of 25~ms each. For each exposure, a frame selection and averaging is subsequently performed, typically using a S/N-based criterion. In the final raw data product
additive biases should be removed, leaving only multiplicative terms to be divided away in the calibration stage. To this end, each sequence of
five science (SCI) exposures is interleaved with similar sequences on well chosen calibrators for which the visibility can be predicted with high precision.
The estimated transfer function contains residual effects from the atmosphere, as well as the interferometer's response to a point source. 

\begin{table*}[t!]
\tabcolsep=10pt
\centering
\caption{Observing log of the AMBER LR data.}             % title of Table
\label{table:obslogAMBER}      % is used to refer this table in the text
% \centering                          % used for centering table
\begin{tabular}{c c c c c c c c}        % centered columns (4 columns)
\hline\hline                 % inserts double horizontal lines
Date & MJD & Calibrator & Stations & DIT & NDIT & FINITO &  \\    % table heading 
&	&	&	& [s]\\
\hline                        % inserts single horizontal line
   2010 Oct 05 $^{(a)}$ & 55474.11 & - & A0-K0-I1 & 0.025 & 1000 & ON  \\   
   2010 Oct 06 & 55475.11 & HD215648$^{(b)}$ & A0-K0-G1 & 0.5 & 120 & ON   \\
   2010 Oct 06 & 55475.16 & HD215648 & A0-K0-G1 & 0.1 & 1000 & ON   \\      % inserting body of the table
\hline                                   %inserts single line
\end{tabular}

\tablefoot{(a) not used; {(b)}: {Diameters (mas): $\theta_{UD}^H$ = 1.08$\pm$0.08, $\theta_{UD}^K$ = 1.08$\pm$0.08}}
\end{table*}

\begin{figure*}[t!]
\centering
   \includegraphics[width=.49\textwidth]{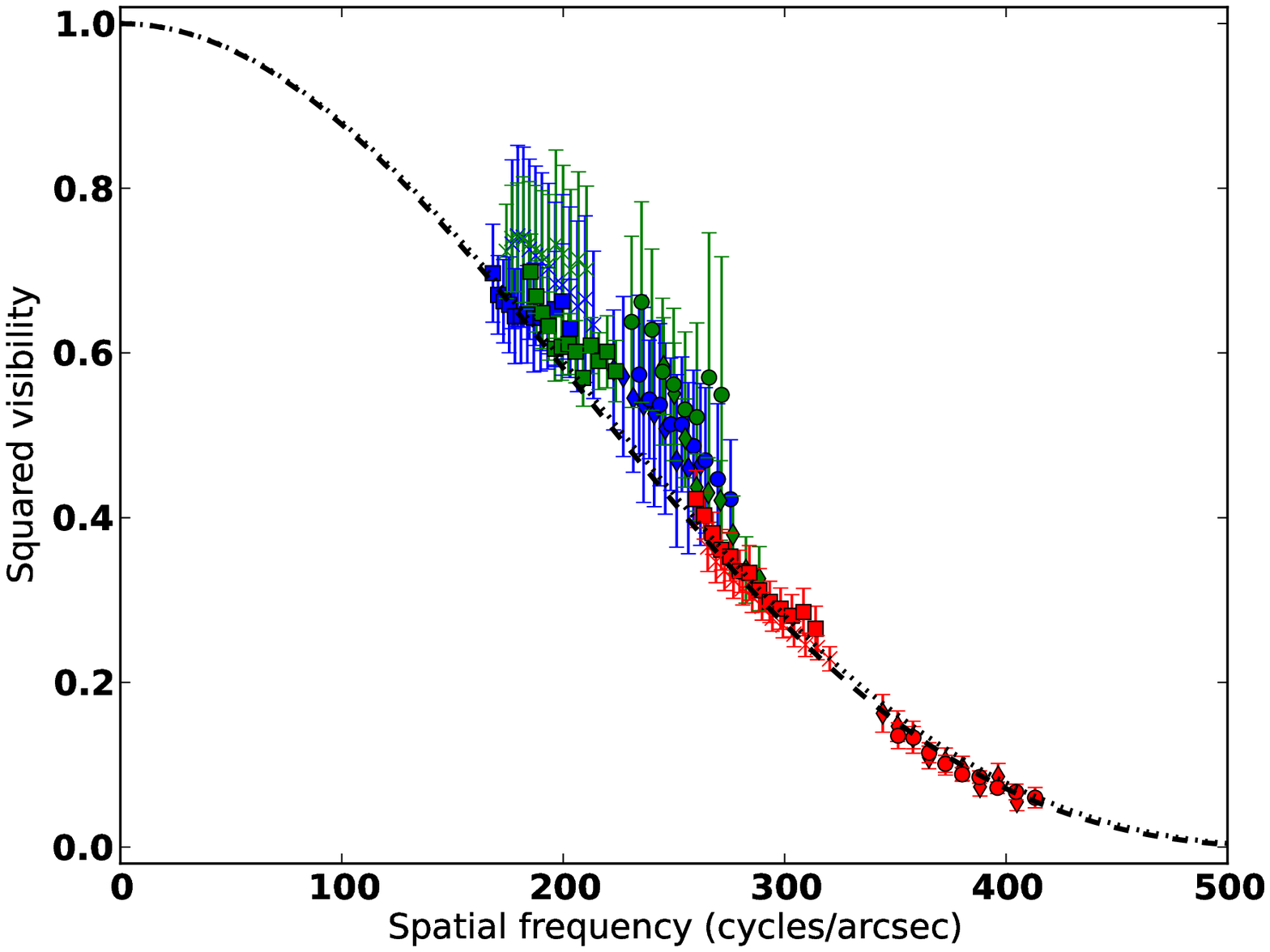} 
   \includegraphics[width=.49\textwidth]{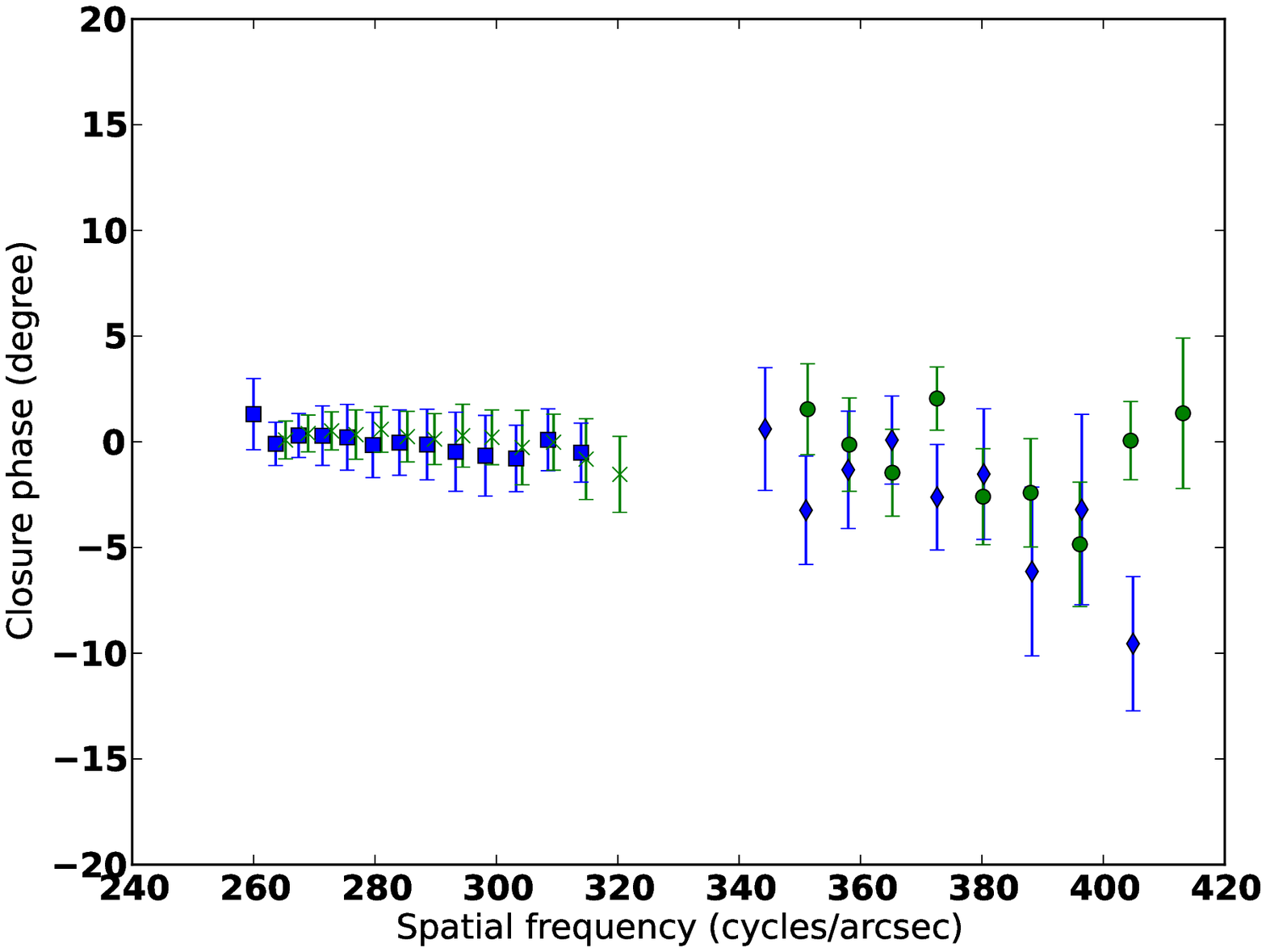}
     \caption{
\new{Left panel: the calibrated squared visibilities of \gpsc. The colour denotes the baseline: K0-G1, A0-G1 and A0-K0 are shown in blue, green and red, respectively. The dashed and dotted lines are the best-fit UD angular diameters in the H and K bands. Different symbols are used to represent the two calibrated exposures in the H (circles and diamonds) and K (squares and crosses) band.
Right panel: the closure phases as a function of the longest spatial frequency in the triplet. The same symbols are used as in the left panel, but the colour now denotes the exposure (blue being the first).}
     \label{figure:interferometrydata} }
\end{figure*} 

During the first night only one sequence of exposures of the science target could be obtained, as well as a sequence on a check star that has the same expected
angular diameter. Since no proper calibrator (CAL) could be observed, these data were not retained in our analysis.
Two sequences of CAL-SCI-CAL were obtained in good observing conditions (seeing $\sim$0.7$\arcsec$) during the second night. With a minimal S/N of $\sim$2 at the longest baseline in the J band, where the VLTI throughput 
is low and the visibility is close to null, the intrinsic data quality is very good. This is also evidenced by the small spread between the subsequent exposures
in each sequence. Despite the good S/N, the J band data were discarded for two reasons. First, the wavelength-calibration is particularly unreliable in the J band. There is no way to do it reliably in the LR \new{observing} mode, so what \textit{amdlib} does is to fit the position of 
the discontinuity between the H and K bands to its expected wavelength value. This works reasonably in the H and K bands, with an estimated systematic 
uncertainty of 2\%, but leads to much larger uncertainties in the J band. Since the wavelength is used in the computation of the P2VM, and determines 
the spatial frequency of the observation, a wrong wavelength table can significantly bias the final result. Second, the spatial average of the phase of the incoming wave front \citep[i.e. piston, cf.][]{piston2001} and its spread is large in the J band. 

The critical step in the AMBER data reduction is the selection and averaging of frames within one exposure. Depending on the performance of FINITO,
the contrast of a frame can be reduced by the residual effect of atmospheric jitter. At the time of our observations no realtime FINITO data were recorded yet, 
so there is no objective way to do the frame selection. Instead, we resorted to the classical approach and selected frames based on piston ($<\,10\,\mu$m) and 
S/N (best 15\%) for the squared visibilities, while the 80\% frames with the best S/N were retained for the closure phases as these are not
sensitive to piston. The used criteria are a trade-off between competing effects: a too large piston will bias the visibilities, while a too 
strong selection criterion on piston will reduce the S/N too much and might bias the visibilities as well. In general, very little frames were removed based on 
piston, indicating that FINITO locked on the fringes properly, except in the H band on the longest baseline. For the latter we note a systematic piston offset from
zero which appears in both science measurements, despite their different detector integration times (DIT). Although the calibrator measurements are affected similarly,
the brightness difference between the science target and the calibrator might result in a residual effect after the calibration. Different selection criteria 
were tried, and we estimate a maximum bias on our derived angular diameter, due to our choice of selection criteria, 
of $\sim$3\% in K and $\sim$6\% in H. 

\begin{figure}[t!]
\centering
   \includegraphics[width=0.5\textwidth]{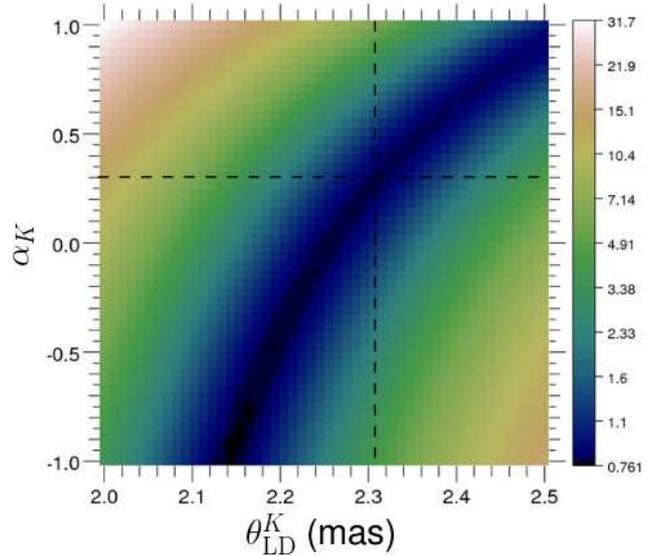} 
     \caption{The $\chi^2_{\rm r}$-map of the limb darkened angular diameter versus the linear limb darkening coefficient.}
     \label{figure:interferometrychi2map}
\end{figure} 

The error on the final calibrated visibilities and closure phases is the quadratic sum of the intrinsic error on the science measurement, estimated from the 
spread between the consecutive exposures, with an error term for the transfer function. The latter is computed as the quadratic sum of the intrinsic 
error on the calibrator measurement with the uncertainty on the calibrator's theoretical visibility, and the spread between the different calibrator measurements. 
The transfer function was very stable on the longest baseline, but less so on the shorter ones; hence the large difference in the errors. 

Special care was taken to estimate the wavelength-dependent uniform disk (UD) angular diameter of the calibrator. 
First, a bolometric diameter is determined from SED-fitting \citep[using the grid-based procedures of][]{2011AADegroote}.
Then the formula from \citet{1974MNRASHanbury}
\begin{equation}
 \frac{\theta_{LD}}{\theta_{UD}} = \sqrt{\frac{1-\alpha_{\lambda}/3}{1-7\alpha_{\lambda}/15}}     \label{eq:Hanbury}
\end{equation}
was used to convert from bolometric to UD diameter. $\alpha_{\lambda}$ is the band-dependent linear limb darkening coefficient for the appropriate stellar parameters, and is taken from the 
catalogue of \citet{ClaretBloemen2011}. Since we use the ATLAS \citep{1993yCatKurucz} plane-parallel atmosphere models, the fitted bolometric 
diameter can be assumed equal to the limb-darkened diameter used in Eq.\,(\ref{eq:Hanbury}) \citep[see e.g.][for a more detailed discussion about this assumption]{2004AAWittkowski}. The derived value (listed in Table~\ref{table:obslogAMBER}) agrees perfectly with the diameter provided by the JMMC's \textit{SearchCal} tool \citep{Bonneau2006}.

The final calibrated squared visibilities and closure phases are graphically depicted in Fig.\,\ref{figure:interferometrydata}. The visibility precision on the longest baseline is high, in contrast to the other baselines. The closure phases are very precise in the K band ($\sim$1-2$^\circ$), and a bit less precise in H ($\sim$3-5$^\circ$), but all are zero within~3$\sigma$.

\subsection{Analysis}
Fitting of simple geometric models to the interferometric data was performed with the LITpro software tool \citep{2008SPIETallonBosc} that is provided by the JMMC.
LITpro does a simple but fast $\chi^2_{\rm r}$-minimisation, that also allows for easy examination of correlations between parameters through the computation of a correlation matrix and $\chi^2$-maps. 
Given the measured closure phases of zero degrees, we only fitted point-symmetric models to the squared visibilities. First we try a uniform disk brightness profile. The H and K band were treated separately to account for the difference in the object's limb darkening and the AMBER systematic errors. 
The best-fit uniform disk curves are shown in Fig.\,\ref{figure:interferometrydata} and have diameters of $\theta_{UD}^H$=2.29$\pm$0.03 and $\theta_{UD}^K$=2.26$\pm$0.04~mas. 
Since LITpro assumes each data point to be independent, while the different AMBER wavelength channels are correlated, we here corrected LITpro's statistical errors using $\sigma = \sqrt{N_{ch}} \sigma_{stat}$.
The corresponding $\chi^2_{\rm r}$ values (with its $\sigma$) are 1.3$\pm$0.2 and 0.8$\pm$0.2 for the H and K bands respectively. 
Although the data do not require a more complex model, it is useful to fit a limb darkened model to check for internal 
consistency and because it is the limb darkened diameter that should be used for comparison with the seismic value. For simplicity we use a linear limb darkening law. 

\new{We note that the visibilities at shorter baselines are systematically above the model fit (Figure\,\ref{figure:interferometrydata}). \new{This is} known to be the result of a calibration with poor transfer function stability (which is mostly systematic across all wavelengths, i.e.  it is to be expected that all visibilities fall on the same side of the model curve). \new{Moreover,} the diameter is constrained mostly by the measurements at the highest spatial frequencies which have a higher resolving power and in this particular case also a more stable transfer function.}

Figure~\ref{figure:interferometrychi2map} shows the $\chi^2_{\rm r}$-map of the limb darkened K band angular diameter versus the linear limb darkening coefficient. 
Since our spatial frequencies only cover the first lobe of the visibility curve, the limb darkening cannot be constrained independently. However, for a 
given limb darkening coefficient we can determine the corresponding diameter. Taking the linear limb darkening coefficients from \citet{ClaretBloemen2011}, using the parameters of \gpsc, which 
are $\alpha_K \in {[0.28,0.31]}$ and $\alpha_H \in {[0.33,0.36]}$, together with Eq.\,(\ref{eq:Hanbury}) and the measured UD diameters, we find limb darkened diameters of $\theta_{LD}^H$=2.4$\pm$0.1\,mas and $\theta_{LD}^K$=2.31$\pm$0.09\,mas, respectively. 
The (conservative) errors are quadratic additions of all mentioned error terms. These limb darkened diameters agree well with the directly 
fitted values for the same value of $\alpha_K$ and $\alpha_H$ (see e.g. Fig.\,\ref{figure:interferometrychi2map}). 
\new{A global fit to the full data set, and assuming the limb darkening coefficients $\alpha_K = 0.295$ and $\alpha_H = 0.345$, results in a final angular diameter of 
$\theta_{LD} = 2.34 \pm 0.10$~mas. 
With a Hipparcos parallax distance of $42.3 \pm 0.3$~pc, this angular diameter translates into a linear radius of $10.6 \pm 0.5$~R$_\odot$.}
%With a Hipparcos parallax distance of 42.3$\pm$0.3\,pc, these angular diameters translate into linear radii of 10.5$\pm$0.4\,R$_\odot$ and 10.7$\pm$0.6\,R$_\odot$, respectively.

% ======================================================

\section{Long periodic variations \label{sec:longPeriodicTrends}}
For both stars, the spectroscopic time series depicted in Fig.\,\ref{fig:curvesRV} reveal  radial velocity variations on time scales longer than 100\,d, which are present in the data sets of all sites separately. Such variations are too low in frequency to be connected to solar-like oscillations.

For the determination of these long periods, only data sets spanning over more than 100 days were used, i.e. \Hermes, \Coralie, and \Hides. The data set from \sophie was excluded from this analysis, as it only covers about a week in time and suffers from strong night to night zero-point shifts. Also the last observing run on \ttau with \Hermes in early January 2011 was excluded from the period determination due to zero-point shifts. We determined the periods by fitting a sine-wave to the variation in the data subset, using \textsc{Period04} \citep{Lenz2005}.

\begin{table}[t!]
\tabcolsep=8pt
\begin{center}
\caption{\label{tab:longPeriodicTrnes}
Long periodic variations found in the campaign stars.}
\begin{tabular}{crrrr}
\hline\hline
Star 		& \multicolumn{1}{c}{Period} 	& \multicolumn{1}{c}{Amplitude} 	& \multicolumn{1}{c}{Phase}	&  \multicolumn{1}{c}{S/N}	\\
 		& \multicolumn{1}{c}{[d]} 		& \multicolumn{1}{c}{[m\,s$^{-1}$]} 	&	& 	\\ \hline
\gpsc 		& 113.0$\pm$0.8 		& 22.8$\pm$0.3 		& 0.789$\pm$0.002 	& 6.4	\\
\hline 
\ttau		& 165$\pm$3		& 9.3$\pm$0.2	& 0.439$\pm$0.004	& 6.4	\\
		& 16.57$\pm$0.05	& 4.7$\pm$0.2	& 0.061$\pm$0.007	& 4.1	\\\hline
\end{tabular}
\end{center}

\tablefoot{The period, amplitude and phase of long periodic variations found in the data set of \gpsc and \ttau. 
The uncertainties were determined through Monte Carlo Simultaneous. The phase is given as fraction of a cycle and with respect to the  time of the first measurement of the data set: 2455415.5 and 2455413.7 for \gpsc and \ttau, respectively. The signal-to-noise ratio S/N was computed for each  \new{oscillation} mode within a frequency box of 0.75 c/d. }
\end{table}

\begin{figure}[t!]
 \begin{center}
 \includegraphics[width=0.5\textwidth, height=35mm]{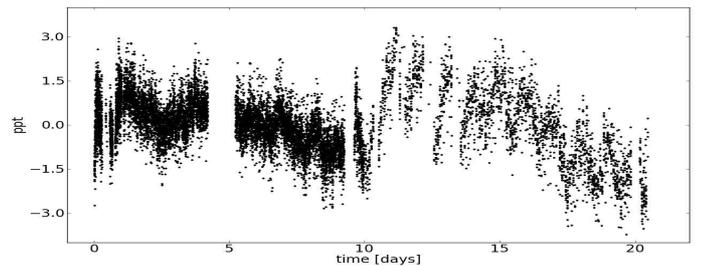}
 \end{center} 
\caption{\label{fig:ttauMost}
MOST photometry of \ttau, deconvolved from thermal trends.}
\end{figure}

\begin{figure*}[t!]
\begin{center}
\includegraphics[width=\textwidth]{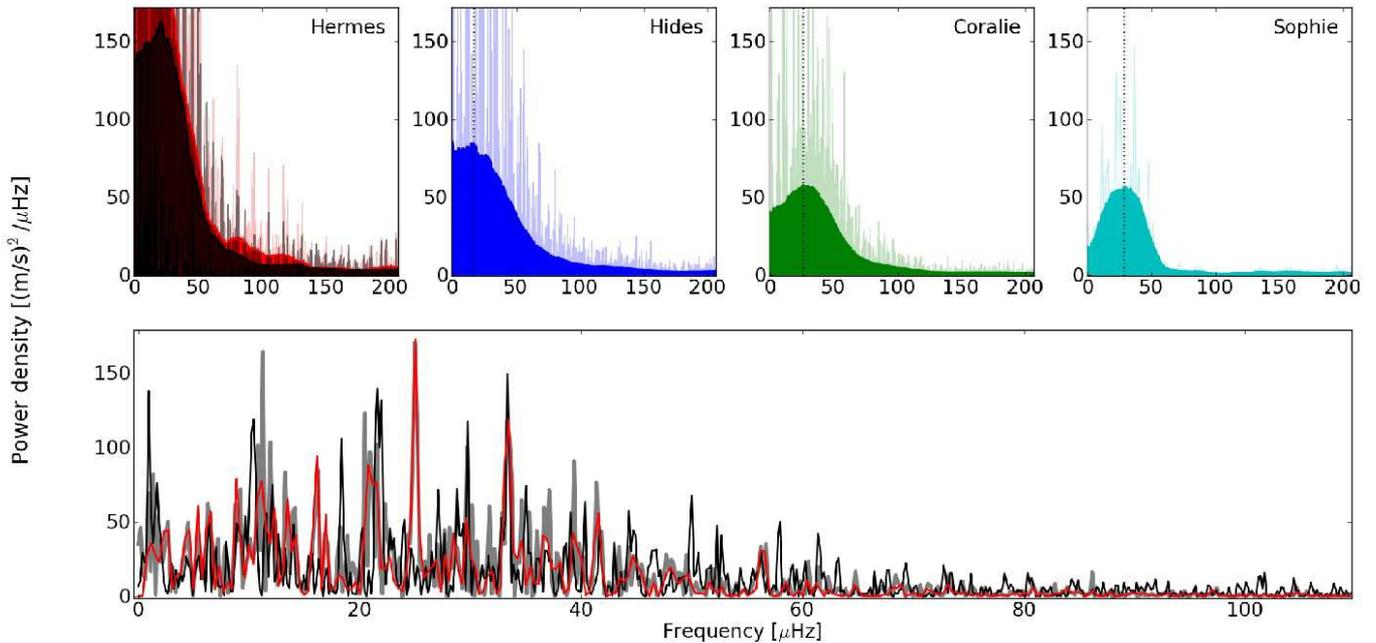}
\end{center}
\caption{\label{fig:gpscSingleSite}
\new{Comparison of the power spectra of \gpsc from the single instruments. 
The top panels show the original and the smoothed signal in the PDS for individual data sets. 
\new{Two PDS are shown in the case of Hermes: the full data set was used for the red PDS and the set corrected for the radial velocity jump at 445 days (Fig.\,\ref{fig:curvesRV}) was used for the black PDS.} The estimate of \num as the highest point of the smoothed PDS is marked as dotted vertical line for each individual data set. The bottom panel shows the PDS of the combined data set, divided into two halves. The first and second half of the data set are shown as grey and black line. The red line shows the power spectrum of the RV-subset with the best coverage (440$<t<$465 days in Fig.\,\ref{fig:curvesRV}). }}
\end{figure*}

For \gpsc we find a variation with a period of 113\,d and a semi-amplitude of 22.8$\pm$0.3\,ms$^{-1}$.  \ttau shows a dominant variation with a period of 165\,d with 9.3$\pm$0.2 ms$^{-1}$ and a smaller variation on the order of 16.57 days (Table\,\ref{tab:longPeriodicTrnes}), which appears to be one-tenth of the main period.
The close up on the radial velocity data of \ttau in the \new{bottom} panel of Fig.\,\ref{fig:rvShorteTimescales} shows that all three data sets are modulated by this intermediate frequency. 

Although we only cover a bit more than 1 cycle, we argue that these variations are real, as they occur in data sets from independent telescopes. Instrumental trends for a given star would not be consistently present with the same time scale and amplitude in both stars in independent data sets. In Fig.\,\ref{fig:curvesRV}, all single-site data sets have been shifted by a constant offset in radial velocity. The variations in radial velocity follow a particular pattern throughout the full time span, independent of the used instrument. We therefore conclude that these trends are intrinsic to the stars.

%\newpage
Estimating the rotation period from the values of the stellar radius and the projected surface rotation velocity in Table\,\ref{tab:literatureValues}, points toward a period of 113 and 138\,days for \gpsc and \ttau, respectively. This is a perfect match to the period of 113$\pm$0.8\,d we found in \gpsc~\new{and suggest that the inclination of the rotation axis of \gpsc is close to 90$^\circ$}. For \ttau the expected value deviates by $\sim$30 days from the 165$\pm$3\,d variation, found from our analysis. \new{\cite{Torres1997} and \cite{Griffin2012} have reported that the inclination of the eccentric binary orbit is probably close to 90 degrees. However, it is questionable if the orbital inclination is identical to the inclination of the rotation axis. For the eccentric system KIC\,5006817, \cite{beck2014a} did not find a good match for between the two axes. The orbital and rotation axis could be inclined by a few degrees. Yet, a lower inclination does not explain this difference.}

In the literature, we find independent proof of the long periodic variability of \ttau. 
The peak-to-peak amplitude we measure for \ttau  is compatible with the radial velocity scatter of $\sim$20\,ms$^{-1}$, found by \cite{Sato2007} in the residuals of the binary model. 
The emission in the Ca\,\textsc{ii}\,H and K lines at 393.4 and 396.9\,nm, respectively shows that \ttau as well as \gpsc are chromospherically active. \new{From about a decade of observations, \cite{Choi1995} found that the Ca \textsc{ii} H and K emission in \ttau varies with a mean period of 140$\pm$18 days, which is in
agreement with our value of 165$\pm$3 days. Given that our value is determined
from only a bit more than one stellar revolution, we conclude that we measured 
the surface rotation of \ttau but that our value could be affected by activity or the short temporal base line.
The found long periods are typical rotation periods for red giants. Indeed, following a different approach, \cite{beck2014a} derived a similar period of 165 days for the surface rotation from forward modelling of the rotational splitting of dipole modes in the red giant component of KIC\,5006817. Although this star a radius of only 6\,R\sun, this seismic result points to the same period regime of surface rotation. Ongoing studies of surface rotation of red giants from solar-like oscillations from \Kepler photometry
point to the same regime of rotation frequencies
and show that there is a  large spread of rotation periods
that depends on the individual rotation history of the
star \citep[e.g.][]{Garcia2014a,McQuillan2014}.}

%From about a decade of observations} \cite{Choi1995} found that in \ttau the Ca\,\textsc{ii}\,H and K emission vary with a \new{mean} period of 140$\pm18$ days, which is in agreement with our value of 165$\pm$3 days \new{which is determined from a bit more than one stellar revolution. We therefore argue that indeed we measure the surface rotation of \ttau, which could be affected by activity or the short temporal base line.}
%\new{These are typical rotation periods for red giants. Following a different approach}, \cite{beck2014a} recently derived a similar period of 165 days for the surface rotation from forward modelling of the rotational splitting \new{of  dipole modes in the red giant component of} KIC\,5006817. \new{Although this star only has 6\,R\sun, the seismically determined rotation period also points into the same period regime of surface rotation. Current and forthcoming studies of surface rotation in solar-like oscillators and red giant stars from Kepler photometry point onto the same regime of rotation frequencies and show that there is a rather large spread of rotation periods that depends on the individual rotation history of the star \citep[e.g.][]{Garcia2014a,McQuillan2014}.}

We note that rotation does not explain the 5\,m\,s$^{-1}$ variation on 16.5 days found for \ttau.  
The available MOST photometry suggests variations on the order of 11 days (Fig.\,\ref{fig:ttauMost}) such that variability with this type of periodicity is confirmed in this independent data set.

\begin{figure*}[t!]
\includegraphics[width=0.49\textwidth]{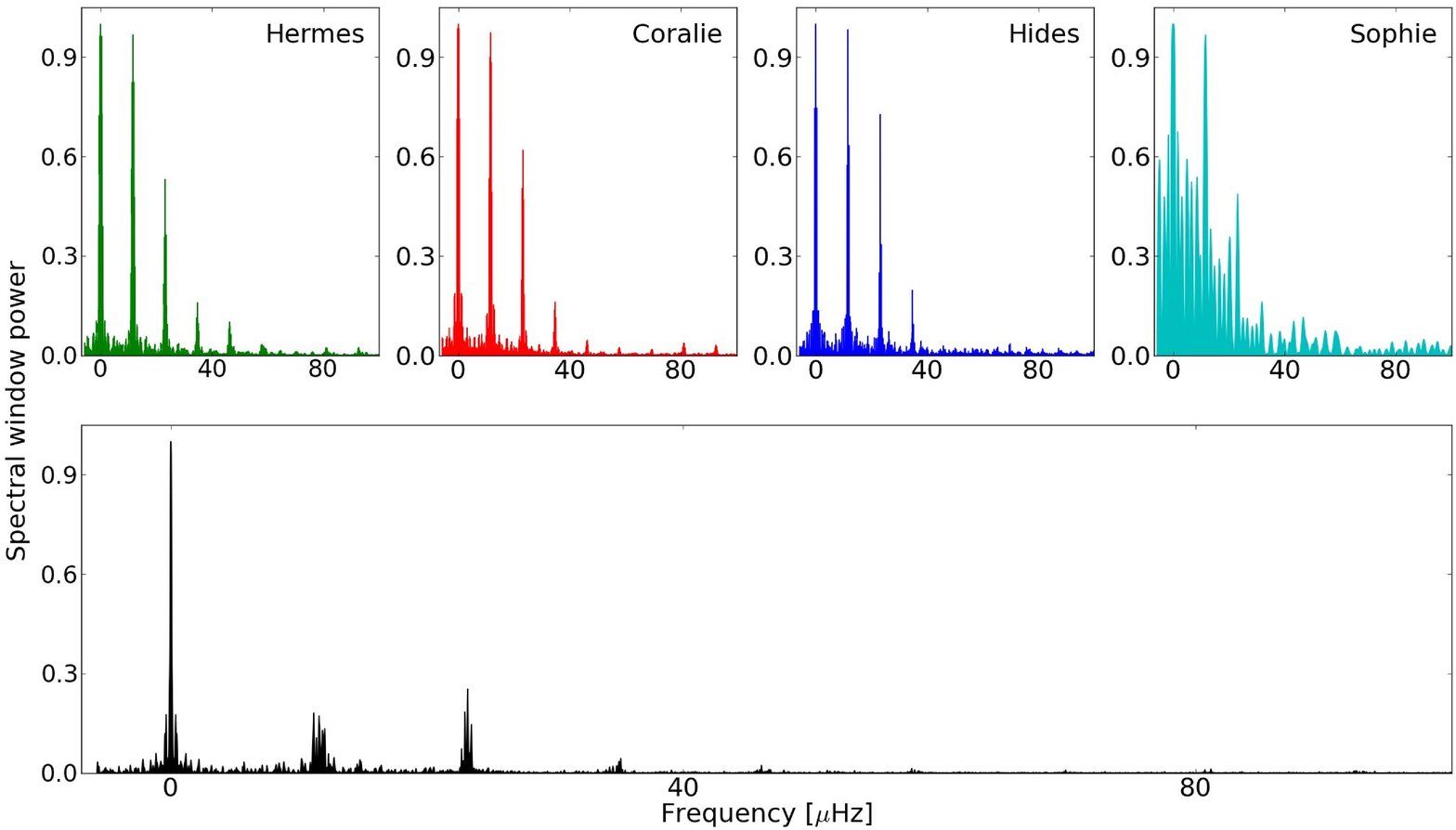}
\includegraphics[width=0.49\textwidth]{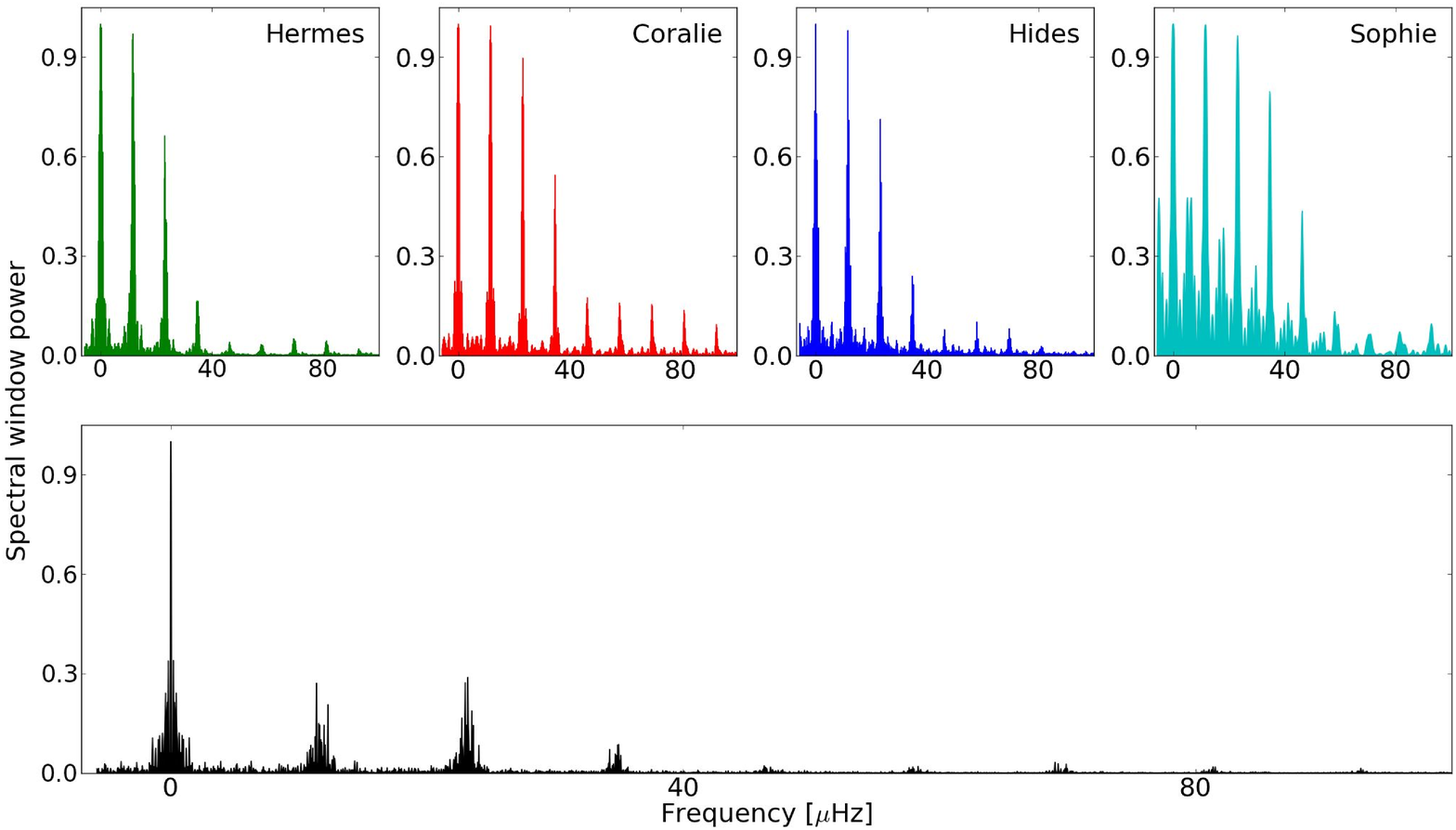}
\caption{\label{fig:spectralWindowTTAU} 
Spectral windows of \gpsc (left panels) and \ttau (right panels).
The upper panels depict the spectral windows of the single-site data.
The spectral window of the combined data set is shown in the lower left panel.\smallskip}
%\end{figure*}
%\begin{figure*}[t!]
\includegraphics[width=\hsize,height=50mm]{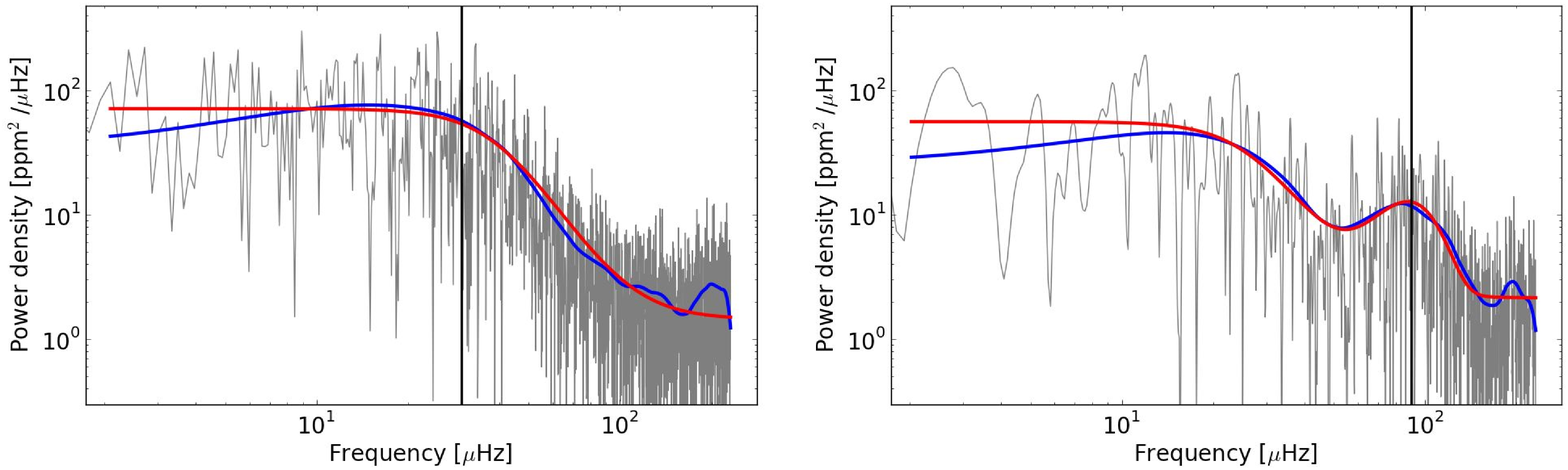}
\caption{\label{fig:harveyLaws}
Power density spectrum of \gpsc (left) and \ttau (right). The original spectrum of the spectroscopic time series is shown in grey. The blue line shows the smoothed PDS. The red line depicts the multi component fit to the original PDS. For \gpsc no Gaussian profile was fitted for \num. The vertical black line shows the adopted value of \num for \gpsc and \ttau of $\sim$32 and 90$\mu$Hz, respectively. For both stars, Fig.\ref{fig:rvShorteTimescales} shows the clear variations on these time scales in the time series. }
\end{figure*}

%\newpage
\section{Solar-like oscillations in \gpsc 
\label{sec:solarlikeOscillationsGpsc}}

The radial velocity curve of \gpsc, depicted in Fig.\,\ref{fig:rvShorteTimescales} and \new{Fig.\,\ref{fig:curvesRV}} (top panel) reveals the clear presence of long and multi periodic radial velocity variations.  Such patterns cannot originate from noise or instrumental trends, as they are consistently present in independent single-site data sets. The key values to describe solar-like oscillations are the frequency of the maximum excess of oscillating power, \num, and the large frequency separation, $\Delta\nu$, between consecutive pressure modes of the same spherical degree $l$.  

From a data set \new{contaminated with zero-point shifts} we cannot determine a precise value for the frequency of maximum oscillation power. A \new{typical} approach would consist of shifting all individual nights to their nightly mean and then determine \num from the combined master spectrum. 
However, \new{from visual inspection of the radial-velocity variations (cf. Fig.\,\ref{fig:rvShorteTimescales}) we estimated the frequency of the oscillations for \gpsc to be on the order of 2 to 3 per day. Therefore, shifting the nightly data sets by their zero point} would act as a high-pass band filter which shifts power to higher values. \new{From comparing the power spectra of zero-point shifted and unshifted data, we find that} the frequency range up to $\sim$30\,$\mu$Hz is severely affected, but also the distribution in the higher frequent part of the PDS is flattened. \new{Such filter would  change the shape of the power excess and mimic a higher \num.}
\new{To avoid unnecessary manipulations of the background signal, we} only prewhitened the intrinsic long periodic variation of 110 days.

\begin{figure*}[t!]
\begin{center}
\includegraphics[width=\hsize]{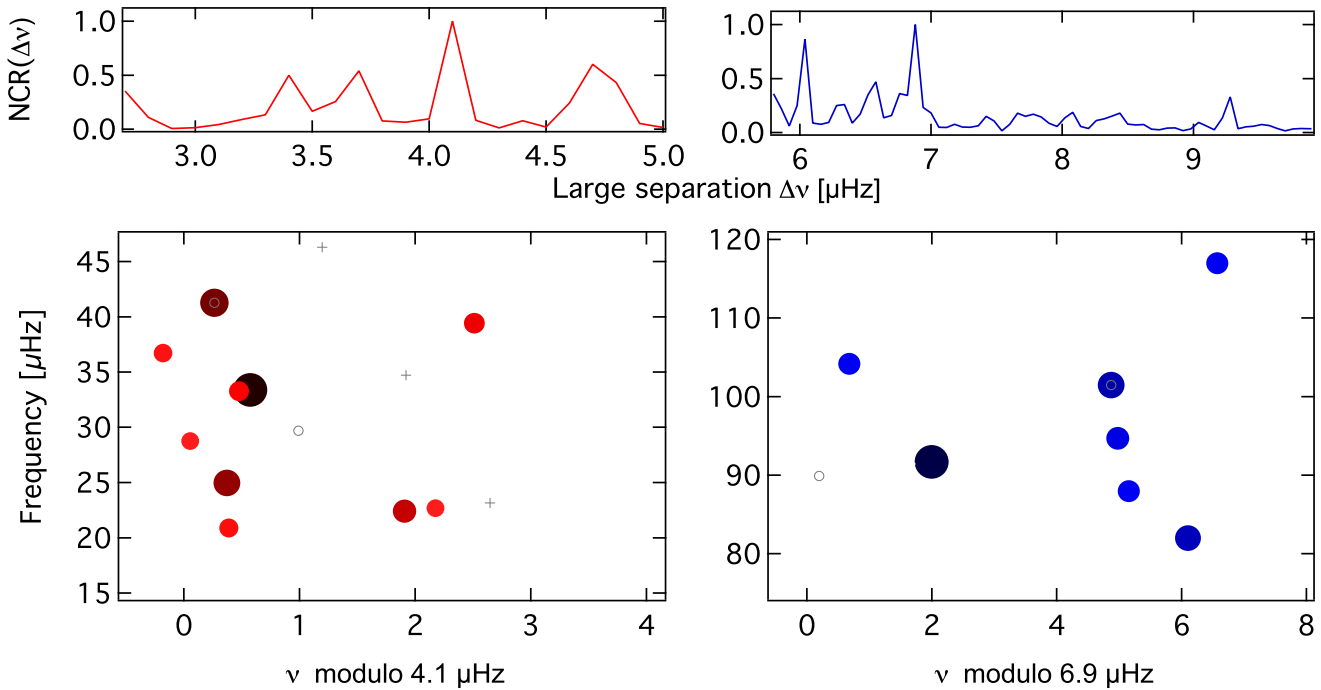}
\end{center}
\caption{\label{fig:combResponse}
Frequency analysis of the combined data set for \gpsc (left) and \ttau (right). The top panels show normalised comb response (NCR) of the 6 strongest frequencies. The \'echelle diagrams (bottom panels) were calculated for the highest peak in the comb response function of \gpsc and \ttau, respectively. Solid dots represent modes with an S/N$\geq$5. Open circles show the original position of modes that were shifted by 11.574\,$\mu$Hz. The crosses in the \'echelle diagram for \gpsc mark the position of multiples of the daily aliases.}
\end{figure*}

Before calculating the power spectrum for the full data set, the data sets for individual instruments were investigated. 
Nearly all single-site data already reveal a good approximation of the frequency of the oscillation power excess \new{(Fig.\,\ref{fig:gpscSingleSite}, top panels)}. \new{The \hermes data contain a source of instrumental noise that is dominated by} long periodic noise and shows a maximum power at very low frequencies. This signal originates from small instabilities of the calibration lamp, which lead to night-to-night shifts. The main shift is visible in the red RV-curve in Fig.\,\ref{fig:curvesRV} between 440 and 460 days as a jump of $\sim$25\,ms$^{-1}$. By correcting for this jump manually, the signal below $\sim$1\,d$^{-1}$ drops \new{(see Fig.\,\ref{fig:gpscSingleSite}, top left panel)}. 
%We therefore could identify the approximate frequency of the power excess \new{in \gpsc} from single-site data.
For further understanding of the night-to-night shifts of the \Hermes spectrograph, we \new{have} initialised a test program on \Hermes~\new{during} winter 2013/14.

From the \new{smoothed power spectrum of the} \new{individual single data sets (Fig.\,\ref{fig:gpscSingleSite}, top panels), and the power spectrum of the combined data set \new{(Fig.\,\ref{fig:gpscSingleSite}, red line in the lower panel)}} we identify the \new{highest signal as the} maximum oscillation power at $\sim$32\,$\mu$Hz.
\new{Variations on this time scale are found from detailed inspection of all independent nightly data sets.} 
Such long-periodic oscillations are expected for red giants higher up on the RGB or red clump stars and are compatible with the position of \gpsc in the Hertzsprung-Russell diagram (HRD).

We subsequently combined all four single-site data sets of \gpsc. The \new{spectral} window of the single-site data sets and the full data set for \gpsc are presented in Fig.\,\ref{fig:spectralWindowTTAU}. The power excess of \gpsc is located exactly at the knee of the power law, describing the long periodic granulation noise (Fig.\,\ref{fig:harveyLaws}). 
\new{Although we see \new{the peak of the power excess} (or aliases of it, shifted by 11.6\,$\mu$Hz) in the individual power spectra (Fig.\,\ref{fig:gpscSingleSite}), \num is hardly visible in the power density spectrum of the combined data set.  We argue that this is also a problem of the zero-point shift between data sets and nights.}
\new{In addition, this region} is contaminated by instrumental noise. 
It is therefore hard to reliably determine the value of \num for \gpsc from a multi-component fit (Fig.\,\ref{fig:harveyLaws}).

%\newpage

In a wide frequency range around the identified maximum of oscillation power \new{(15-50\,$\mu$Hz)}, all significant frequencies (i.e., S/N $\geq$ 5) were prewhitened, using  the ISWF (iterative sine-wave fitting) method. 
The S/N compares the height of a mode to the mean noise in a selected frequency range. For \gpsc it was determined between 50 and 70\,$\mu$Hz. Indeed, we find the highest peaks around $\sim$32\,$\mu$Hz. 

The large frequency separation was then computed from the generalised comb-response (CR) function \citep{Bonanno2008, corsaro2012b}, based on the six frequencies with the highest power. \new{The extracted frequencies are available in the online material.} With the CR function, we scanned a wide range of potential values of \dnu, which includes a mass range that generously frames the mass estimates for the star in the literature. For \gpsc  the chosen range corresponds to $\sim$0.5 to $\sim$5\,M\sun\ (Fig.\,\ref{fig:combResponse}). The values of \dnu, suggested by the highest peaks were then tested with \'echelle diagrams.

The CR-function, calculated from the extracted peaks, exhibits a clear peak at $\Delta\nu$=4.1$\pm$0.1\,$\mu$Hz (Fig\,\ref{fig:combResponse}, left upper panel). 
The uncertainty originates from a Gaussian fit to the main peak and given that we are dealing with ground-based and gapped data can only be taken as a lower limit. 
\new{In addition, we tested different step-sizes of the CR function (up to 0.02\,$\mu$Hz) but did not find a different result for the value and its uncertainty.}
This value remains stable up to the inclusion of the seventh highest out of the ten most significant frequencies. This is not surprising, because some of them are likely to be mixed modes, which do not follow the expected regular frequency pattern of pure p-modes. Using this value and the set of significant frequencies, we constructed an \'echelle diagram for this star, showing \new{one ridge} (Fig.\,\ref{fig:combResponse}). 
Most of the strongest modes are lining up on the left side of the diagram (i.e. $\nu$\,modulo\,$\Delta\nu\leq1$, hereafter termed \textit{\'echelle phase}). \new{We note that the \'echelle phase is closely related through the modulo operation to the parameter $\varepsilon$ from the asymptotic frequency relation \new{\citep{Aerts2010}.}} One of the extracted frequencies is likely to be shifted by daily aliases.
When we correct the strong mode at 29.7\,$\mu$Hz for the daily alias, it also falls in the ridge of strong modes. The original and the shifted position of a mode in the \'echelle diagram are marked by open circles. This structure could be the ridge of radial and quadrupole modes. 
\new{We can compare our identification to the $\varepsilon$-values found in large samples of red giant stars, observed with the \textit{Kepler} space telescope \citep[e.g.][]{Huber2010, corsaro2012}. For these stars we have a solid mode identification from uninterrupted data. 
Indeed, the study of \cite{Huber2010} shows that for a star with a \dnu of $\sim$4\,$\mu$Hz, one expects a value of 
\new{$\varepsilon$$\simeq$1} for radial modes suggesting that we correctly identified the radial modes.}
The \new{remaining three}  extracted frequencies (at \'echelle phase of $\sim$2) may originate from mixed dipole modes, \new{which is in agreement with \new{$\varepsilon$$\simeq$1.5} for dipole modes}. A comb-like pattern with a large separation of 4.1\,$\mu$Hz cannot be explained by daily aliasing, as this value of \dnu is not a multiple of the daily alias (11.574\,$\mu$Hz). For comparison the position of the daily alias frequencies are depicted as crosses in the \'echelle diagrams. 

Following the scaling relations described by \cite{kjeldsen1995}, \new{using the revised reference values of \cite{Chaplin2011}}, we estimate \gpsc to be a star of M$\simeq$0.9\,M$_\odot$ and R$\simeq$10\,R\sun. The stellar radius from the scaling relations is in good agreement with the independent radius of \new{10.6$\pm$0.5\,R\sun} from interferometric observations (see Section\,\ref{sec:interferometry}). 
The mass obtained from both scaling relations are below the mass derived from spectroscopic calibrations by \cite{Luck2007} to be  1.87\,M$_\odot$. No uncertainty was given for this value.
\new{These values are likely affected by the lower constraining level of \num, if compared to the case of \ttau (see Sect. 6). However, from the \'echelle diagram (Fig.\,\ref{fig:combResponse}) the value of the large separation appears convincing. By using the value of the large separation found from our oscillation spectrum of \gpsc, and the independent radius from interferometry for the scaling relation for the stellar mass,
\begin{equation}
\label{eq:comparingDnu}
\frac{M}{M_\odot} = \left( \frac{\Delta\nu}{\Delta\nu_\odot}\right)^2 \cdot \left(\frac{R}{R_\odot}\right)^3
%\frac{\Delta\nu}{\Delta\nu_\odot} \simeq \sqrt{\frac{\rho}{\rho_\odot}},
%\frac{M}{M_\odot} =  \frac{\rho}{\rho_\odot} \cdot (\frac{R}{R_\odot})^3
\end{equation} we can derive the mass of the star without the use of \num. This approach leads to a mass of 1.1$\pm$0.2\,M\sun for \gpsc, which is in agreement with the mass found from the scaling relations.} We therefore adopt 32\,$\mu$Hz and 4.1\,$\mu$Hz as \new{good} estimates of the $\nu_{\rm max}$ and $\Delta\nu$ in \gpsc, respectively. 

%\cite{Mosser2013} suggested that the Sun is not the optimal reference point for the scaling relations of red giants. By adapting the scaling relations to their suggested new $\nu_{\rm max}^{\rm ref}$ and $\Delta\nu^{\rm ref}$, we find \gpsc slightly more massive and larger: 1.1\,M$_\odot$ and 11\,R$_\odot$. 
%Given that these values are derived from approximations of the seismic parameters the difference is negligible.

\section{Solar-like oscillations in \ttau
\label{sec:solarlikeOscillationsTtau}}
For \ttau, we have two independent data sets, that we can analyse for the signature of oscillations. Besides our spectroscopic campaign, covering 190-days, we also have photometric data from the MOST satellite.

\subsection{Spectroscopic campaign}
The data sets of \ttau reveal a clear signature of the power excess at 90$\pm$0.2\,$\mu$Hz. The value of \num was obtained from the fit of a power law plus a Gaussian profile and a constant noise background to the power density spectrum  (Fig.\,\ref{fig:harveyLaws}). This value is a formal error from the fit and we note that this is unrealistically small.

To further improve the data, we corrected for zero-point shifts. As visible from Fig.\,\ref{fig:curvesRV} \Sophie data suffer from severe night to night jumps. Also the last data from \Hermes contain sudden jumps of up 50\,ms$^{-1}$ (at $t>$560\,d in Fig.\,\ref{fig:curvesRV}). A visual inspection of the simultaneous ThArNe frames showed that these jumps originate from instabilities of the calibration lamp. Problematic frames were identified and deleted from the data set. The power excess of \ttau is located at frequencies high enough that allow shifting the individual nights to their mean without manipulating the frequency region of \num. Therefore we subdivided the time string into individual nights and corrected them for the nightly mean. The frequency analysis was then performed at this time series of residuals. In total, we find \new{7} frequencies with an S/N$\geq$5 in the frequency range around \num. \new{The extracted frequencies are available  in the online material.} \new{From comparison with the MOST power spectrum, we find that the strong mode at 89.9\,$\mu$Hz falls on top of the ridge structure when shifting by +1\,d$^{-1}$.}
 As we used the shifted data set \new{for the prewhitening analysis}, we could not use the background determined from the fit of the power-laws but derived the mean noise from the frequency range between 140 and 170\,$\mu$Hz

The comb response function of \ttau (Fig.\,\ref{fig:combResponse}, top right) is complex as it exhibits three peaks in the range of 6\,$\mu$Hz~$<\Delta\nu_{\rm CR}<$~7\,$\mu$Hz. \new{The CR-function} remains stable at this \new{frequency range}, independent of the number of significant \new{oscillation peaks} included \new{and the resolution for which the CR-function is calculated}. From the \'echelle diagram we can exclude $\Delta\nu_{\rm CR}$$\simeq$\,6$\mu$Hz, as by plotting all significant frequencies in an \'echelle diagram, inclined ridges form and it is close to half of the daily aliasing. This indicates a wrong trial value for \dnu.
\new{Peaks above $\sim$8\,$\mu$Hz} can be excluded, as \new{they} suggest a mass \new{below} $\sim$1.5\,M\sun and \new{lead to a final radius ($\lesssim$8\,R\sun) which is $\sim$30\,\% smaller than the interferometric radius 11.7$\pm$0.2\,R$_\odot$, determined by \cite{Boyajian2009}. We therefore exclude these values from the possible solutions.}

\begin{figure*}[t!]
\centering
\includegraphics[width=0.99\textwidth]{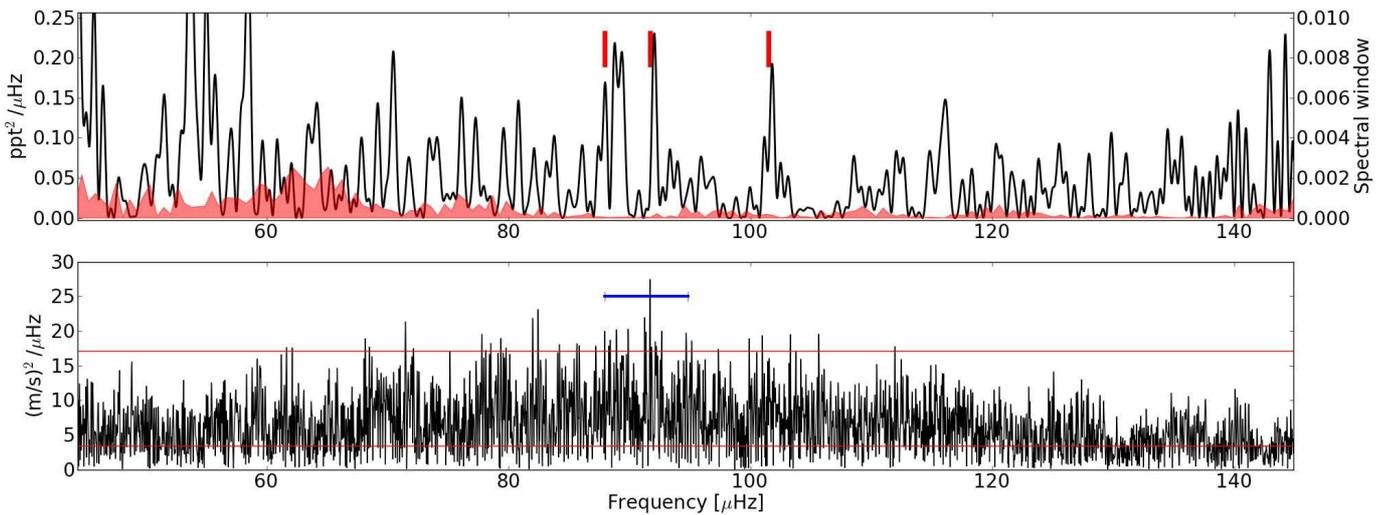}
\caption{\label{fig:mostTTAU}
Comparison of the power spectra of the photometric and spectroscopic PDS of \ttau. 
The top panel shows the power spectrum and the spectral window of the MOST data set as black line and red shaded region, respectively. \new{The spectral window depicts the position of the orbital aliases and reveals that the range around \num is nearly free of contaminations. } The three main peaks in the region around 90\,$\mu$Hz in the MOST spectrum, correspond to frequencies also detected with the ground-based campaign (red dashes). The bottom panel shows the unprewhitened power spectrum of the combined spectroscopic campaign data. The blue horizontal line indicates an average large frequency separation found for \ttau. The lower and upper horizontal red line indicates the noise level and 5 times the noise level, respectively.} \end{figure*}

The highest peak in the CR-function indicates a large separation of 6.9$\pm$0.2$\mu$Hz. Again we note, that the error is likely to be an underestimate. The \'echelle diagram of \ttau shows 8 frequencies with an S/N$\geq$5 \new{folded with  \dnu = 6.9\,$\mu$Hz (Fig.\,\ref{fig:combResponse}, bottom right)}. \new{From the scaling relations of \cite{kjeldsen1995}, we }obtain from mass of 2.7$\pm$0.3\,M\sun \new{and a} radius of 10.2$\pm$0.6\,R\sun for a given \dnu=6.9$\pm$0.2\,$\mu$Hz.
\new{For this value, the mass of the red giant primary of the \ttau-system is well in agreement with the} mass estimate provided by the orbital parameters and the Hipparcos dynamical parallax.
\new{\cite{Torres1997} reported a mass of 2.9$\pm$0.9\,M\sun for the primary component, while using the parallax obtained from Hipparcos data} \cite{Lebreton2001} derived a mass of 2.8$\pm$0.5\,M\sun~and 1.28$\pm$0.1\,M\sun~for the primary and secondary of $\theta^1$\,Tau, respectively. \new{All those values are substantially more massive than 2.3\,M\sun of the turn off point in the Hyades as must be the case \citep{Perryman1998}.}

%\new{ This value of the} \new{large separation leads to a stellar radius that is smaller than the radius reported from interferometric observations.} 
%\newpage
\new{While this value of 6.9\,$\mu$Hz for the large frequency separation leads to a stellar radius that is slightly smaller than the interferometrically determined
radius reported by \citet{Boyajian2009}, there are indications that this interferometric radius is somewhat overestimated (by a few percent), to the extent that the disagreement is not significant. First, the diameters derived by \citet{Boyajian2009} for two other HyadesÕ giants, $\delta^1$\,Tau
and $\epsilon$\,Tau, are larger than other interferometric radii reported in the literature for these giants by 2-3\,$\sigma$, and it is thus plausible that also their diameter for \ttau is too large. Since \citet{Boyajian2009} were limited to a single baseline, calibrator, and observing night, the calibration may have suffered undetected systematics.  Second, this reasoning is further corroborated by the disagreement between the spectroscopic effective temperature from  
\cite{hekker2007} and the T$_{\rm eff}$ derived by \citet{Boyajian2009}, with the latter being 200\,K lower than the former. This also suggests that the diameter derived by \citet{Boyajian2009} is slightly too large, as a T$_{\rm eff}$  derived from a bolometric flux and a radius (the method used by \citet{Boyajian2009}) scales inversely with assumed radius.  
\new{For a more general discussion concerning possible biases in interferometric diameters, we refer to 
\citet{Casagrande2014}.}
We therefore refrain from using this interferometric diameter as an argument against a seismic mass corresponding to 6.9\,$\mu$Hz frequency spacing, and suggest the need for a new interferometric diameter determination, based on multiple baselines and calibrators. Given the required time frame of such an observational campaign, it is outside the scope of the current paper. }

%\newpage
\subsection{Oscillation in MOST space photometry \label{sec:MOST}}
Independent confirmation of the intrinsic variability and oscillations in \ttau is provided by the MOST space photometry. Figure\,\ref{fig:mostTTAU} compares the power spectra of the MOST data set and the spectroscopic campaign. The variability of \ttau is on the detection limit of the satellite. The spectral window of the data set, depicted as red shaded surface, shows that the power spectrum of the photometric measurements does not follow the shape of the spectral window and that the contamination by aliasing in this region is minor. A typical power density diagram can not be drawn, as the spectrum is dominated by the orbital aliasing. The main orbital alias frequencies are located well above this range at 164.35\,$\mu$Hz (14.2 d$^{-1}$). 

The strongest signal found in the photometric data is around 90\,$\mu$Hz and located in a region without contamination. We find that the three highest peaks from MOST photometry coincide with significant frequencies from the spectroscopic campaign data set. 
These frequencies are marked with red dashes in Fig.\,\ref{fig:mostTTAU}. This independently confirms our detection of oscillation power in this region from the spectroscopic campaign data.

\section{Conclusions}
In this paper, we report on the spectroscopic multisite campaign performed in 2010, dedicated to the single field star \gpsc and on \ttau, the primary component of an eccentric binary system in the Hyades open cluster. This campaign was supported by interferometric observations from the VLTI and space data from the MOST satellite.

We found clear evidence of long and multi periodic variations in the radial velocity data set of \gpsc.
No precise value of \num could be deduced due to contamination by granulation noise, and instrumental trends. However, we see clear indications, that \num is located around $\sim$32\,$\mu$Hz. This value is typical for red clump stars and agrees with the position of \gpsc in the HRD. By analysing the frequency range around the adopted \num, we find a large separation of 4.1$\pm$0.1$\mu$Hz from the CR function. 
This value also leads to a convincing \'echelle diagram. The stellar radius from scaling relations shows good agreement with the radius from interferometry. We therefore estimate the mass of \gpsc to be $\sim$1\,M$_\odot$.

\ttau exhibits a clear power excess at 90\,$\mu$Hz in the spectroscopic data sets. Oscillations in this range and the dominant peaks are independently confirmed from MOST space photometry. According to the CR-function analysis a large separation of either 6.9\,$\mu$Hz is preferred. \new{Under the reasonable assumption that the interferometric radius of \ttau \citep{Boyajian2009} is overestimated, we find a rather good agreement between this value and the radius from the scaling relations. } 
\new{A new set of observations with the current instruments and the longest baselines on the CHARA array or similar instrumentation would allow a more reliable measurement of the radius of this star.}
The mass of \ttau is 2.7$\pm$0.3\,M\sun.

\ttau is of special interest due to its membership in the Hyades. Since 2010, the radial velocity of the binary system \ttau~\new{has been} monitored with \Hermes. The binary is approaching its periastron passage early 2014. We hope to be able to perform a spectral disentangling analysis from these data.

In addition, we note that the value of the large separation for both stars comes from an attempt to measure the \new{actual mean} \dnu of the stars, because we only relied on the computation of the comb response function. Further investigations are needed to confirm these values. Unfortunately, the quality of the PDS does not allow to discriminate between the evolutionary stages from mixed mode period spacings. However, the frequency of the maximum oscillation power excess is typical for red clump stars.

Besides the solar-like oscillations, we also detected long periodic modulations of the radial velocities, which occur on the order of 113 days and 165 days for \gpsc and \ttau, respectively. These time scales are in good agreement with the estimates of the rotation period from the stellar radius and the measured $v\sin i$. 

From this study we learned that the common approach to correct for zero-point variations by shifting the nightly data set of individual telescopes will suppress frequency information and shift \num to higher frequencies for power excesses below $\sim$40\,$\mu$Hz. 
We note, that also stars with \dnu$\simeq$11.6\,$\mu$Hz, are very challenging from ground, as their large separation \dnu is identical to the 1-day aliasing in ground-based data. \new{Very good} spectral windows are needed to understand the oscillation spectra of such stars. Also, the interplay of solar-like oscillations and surface rotation requires a dense sampling and consistent data sets in order to disentangle the two effects. 
This can only be achieved with very dense telescope networks.

Finally, we note that in times of stunning asteroseismic results from space photometry, often the benefits of ground-based spectroscopy are forgotten. It is gradually becoming harder to successfully propose for sufficient telescope time for campaigns like the one described here. A promising future ground-based project is the \textit{Stellar Observations Network Group, SONG}\footnote{http://song.phys.au.dk/}, which is a Danish led project to construct a global network of small 1\,m telescopes around the globe \citep{Song2011}. One of the main goal of the project is to obtain time series of radial velocity measurements with a high duty cycle, similar to our campaign. This will therefore allow us to investigate in detail the asteroseismic properties of the stars.

\begin{acknowledgements}
The research leading to these results has received funding from the European Research Council under the European Community's Seventh Framework Programme (FP7/2007--2013)/ERC grant agreement n$^\circ$227224 (PROSPERITY). This research is (partially) funded by the Research Council of the KU Leuven under grant agreement GOA/2013/012. EK is supported by Grants-in-Aid for Scientific Research 20540240 from the Japan Society for the Promotion of Science (JSPS). 
KZ receives a Pegasus Marie Curie Fellowship of the Research Foundation Flanders (FWO).
MY, HI, and BS is supported by Grants-in-Aid for Scientific Research 18340055 from JSPS.
JDR, and E.C. acknowledge the support of the FWO-Flanders under project O6260 - G.0728.11. TK also acknowledges financial support from the Austrian Science Fund (FWF P23608). SB is supported by the Foundation for Fundamental Research on Matter (FOM), which is part of the Netherlands Organisation for Scientific Research (NWO). VSS is an Aspirant PhD fellow at the Fonds voor Wetenschappelijk Onderzoek (FWO), Vlaanderen.
\new{EM is a beneficiary of a mobility grant from the Belgian Federal Science Policy Office co-funded by the Marie Curie Actions
FP7-PEOPLE-COFUND-2008 n¡246540 MOBEL GRANT from the European Commission.}
The ground-based observations are based on spectroscopy made with the Mercator Telescope, operated on the island of La Palma by the Flemish Community, at the Spanish Observatorio del Roque de los Muchachos of the Instituto de Astrof'sica de Canarias. 
We are grateful to the Geneva ObservatoryÕs technical staff for maintaining the 1.2\,m Euler Swiss telescope and the \Coralie spectrograph.
We are grateful to all staff members at OAO, National
Astronomical Observatory of Japan (NAOJ) for their continuous
and various support throughout our project with \Hides.
We thank the technical team at Haute-Provence Observatory for their support with the \Sophie instrument and the 1.93-m OHP telescope. \new{We thank the referee for constructive comments and stimulating discussions.}
\end{acknowledgements}

\bibliographystyle{aa}
\bibliography{bibliography}

\begin{thebibliography}{81}
\expandafter\ifx\csname natexlab\endcsname\relax\def\natexlab#1{#1}\fi

\bibitem[{{Aerts} {et~al.}(2010){Aerts}, {Christensen-Dalsgaard}, \&
  {Kurtz}}]{Aerts2010}
{Aerts}, C., {Christensen-Dalsgaard}, J., \& {Kurtz}, D.~W. 2010,
  {Asteroseismology, Springer Science+Business Media B.V.}

\bibitem[{{Ando} \& {Osaki}(1975)}]{Ando1975}
{Ando}, H. \& {Osaki}, Y. 1975, \pasj, 27, 581

\bibitem[{{Arentoft} {et~al.}(2008){Arentoft}, {Kjeldsen}, {Bedding}, {Bazot},
  {Christensen-Dalsgaard}, {Dall}, {Karoff}, {Carrier}, {Eggenberger},
  {Sosnowska}, {Wittenmyer}, {Endl}, {Metcalfe}, {Hekker}, {Reffert}, {Butler},
  {Bruntt}, {Kiss}, {O'Toole}, {Kambe}, {Ando}, {Izumiura}, {Sato}, {Hartmann},
  {Hatzes}, {Bouchy}, {Mosser}, {Appourchaux}, {Barban}, {Berthomieu},
  {Garcia}, {Michel}, {Provost}, {Turck-Chi{\`e}ze}, {Marti{\'c}}, {Lebrun},
  {Schmitt}, {Bertaux}, {Bonanno}, {Benatti}, {Claudi}, {Cosentino}, {Leccia},
  {Frandsen}, {Brogaard}, {Glowienka}, {Grundahl}, \&
  {Stempels}}]{arentoft2008}
{Arentoft}, T., {Kjeldsen}, H., {Bedding}, T.~R., {et~al.} 2008, ApJ, 687, 1180

\bibitem[{{Arentoft} {et~al.}(2014){Arentoft}, {Tingley},
  {Christensen-Dalsgaard}, {Kjeldsen}, {White}, \& {Grundahl}}]{Arentoft2014}
{Arentoft}, T., {Tingley}, B., {Christensen-Dalsgaard}, J., {et~al.} 2014,
  \mnras, 437, 1318

\bibitem[{{Auriere} {et~al.}(2014){Auriere}, {Konstantinova-Antova},
  {Charbonnel}, \& {et}}]{Auriere2014}
{Auriere}, M., {Konstantinova-Antova}, R., {Charbonnel}, C., \& {et}, a. 2014,
  in preparation

\bibitem[{{Beck} {et~al.}(2011){Beck}, {Bedding}, {Mosser}, {Stello}, {Garcia},
  {Kallinger}, {Hekker}, {Elsworth}, {Frandsen}, {Carrier}, {De Ridder},
  {Aerts}, {White}, {Huber}, {Dupret}, {Montalb{\'a}n}, {Miglio}, {Noels},
  {Chaplin}, {Kjeldsen}, {Christensen-Dalsgaard}, {Gilliland}, {Brown},
  {Kawaler}, {Mathur}, \& {Jenkins}}]{beck2011}
{Beck}, P.~G., {Bedding}, T.~R., {Mosser}, B., {et~al.} 2011, Science, 332, 205

\bibitem[{{Beck} {et~al.}(2010){Beck}, {Carrier}, \& {Aerts}}]{beck2010}
{Beck}, P.~G., {Carrier}, F., \& {Aerts}, C. 2010, AN, 331, P32

\bibitem[{{Beck} {et~al.}(2014){Beck}, {Hambleton}, {Vos}, {Kallinger},
  {Bloemen}, {Tkachenko}, {Garc{\'{\i}}a}, {{\O}stensen}, {Aerts}, {Kurtz}, {De
  Ridder}, {Hekker}, {Pavlovski}, {Mathur}, {De Smedt}, {Derekas}, {Corsaro},
  {Mosser}, {Van Winckel}, {Huber}, {Degroote}, {Davies}, {Pr{\v s}a},
  {Debosscher}, {Elsworth}, {Nemeth}, {Siess}, {Schmid}, {P{\'a}pics}, {de
  Vries}, {van Marle}, {Marcos-Arenal}, \& {Lobel}}]{beck2014a}
{Beck}, P.~G., {Hambleton}, K., {Vos}, J., {et~al.} 2014, \aap, 564, A36

\bibitem[{{Beck} {et~al.}(2012){Beck}, {Montalban}, {Kallinger}, {De Ridder},
  {Aerts}, {Garc{\'{\i}}a}, {Hekker}, {Dupret}, {Mosser}, {Eggenberger},
  {Stello}, {Elsworth}, {Frandsen}, {Carrier}, {Hillen}, {Gruberbauer},
  {Christensen-Dalsgaard}, {Miglio}, {Valentini}, {Bedding}, {Kjeldsen},
  {Girouard}, {Hall}, \& {Ibrahim}}]{beck2012}
{Beck}, P.~G., {Montalban}, J., {Kallinger}, T., {et~al.} 2012, Nature, 481, 55

\bibitem[{{Bedding} {et~al.}(2007){Bedding}, {Kjeldsen}, {Arentoft}, {Bouchy},
  {Brandbyge}, {Brewer}, {Butler}, {Christensen-Dalsgaard}, {Dall}, {Frandsen},
  {Karoff}, {Kiss}, {Monteiro}, {Pijpers}, {Teixeira}, {Tinney}, {Baldry},
  {Carrier}, \& {O'Toole}}]{Bedding2007}
{Bedding}, T.~R., {Kjeldsen}, H., {Arentoft}, T., {et~al.} 2007, \apj, 663,
  1315

\bibitem[{{Bedding} {et~al.}(2011){Bedding}, {Mosser}, {Huber},
  {Montalb{\'a}n}, {Beck}, {Christensen-Dalsgaard}, {Elsworth},
  {Garc{\'{\i}}a}, {Miglio}, {Stello}, {White}, {De Ridder}, {Hekker}, {Aerts},
  {Barban}, {Belkacem}, {Broomhall}, {Brown}, {Buzasi}, {Carrier}, {Chaplin},
  {di Mauro}, {Dupret}, {Frandsen}, {Gilliland}, {Goupil}, {Jenkins},
  {Kallinger}, {Kawaler}, {Kjeldsen}, {Mathur}, {Noels}, {Aguirre}, \&
  {Ventura}}]{bedding2011}
{Bedding}, T.~R., {Mosser}, B., {Huber}, D., {et~al.} 2011, Nature, 471, 608

\bibitem[{{Bonanno} {et~al.}(2008){Bonanno}, {Benatti}, {Claudi}, {Desidera},
  {Gratton}, {Leccia}, \& {Patern{\`o}}}]{Bonanno2008}
{Bonanno}, A., {Benatti}, S., {Claudi}, R., {et~al.} 2008, \memsai, 79, 639

\bibitem[{{Bonneau} {et~al.}(2006){Bonneau}, {Clausse}, {Delfosse}, {Mourard},
  {Cetre}, {Chelli}, {Cruzal{\`e}bes}, {Duvert}, \& {Zins}}]{Bonneau2006}
{Bonneau}, D., {Clausse}, J.-M., {Delfosse}, X., {et~al.} 2006, \aap, 456, 789

\bibitem[{{Bouchy} {et~al.}(2009){Bouchy}, {H{\'e}brard}, {Udry}, {Delfosse},
  {Boisse}, {Desort}, {Bonfils}, {Eggenberger}, {Ehrenreich}, {Forveille},
  {Lagrange}, {Le Coroller}, {Lovis}, {Moutou}, {Pepe}, {Perrier}, {Pont},
  {Queloz}, {Santos}, {S{\'e}gransan}, \& {Vidal-Madjar}}]{Bouchy2009}
{Bouchy}, F., {H{\'e}brard}, G., {Udry}, S., {et~al.} 2009, \aap, 505, 853

\bibitem[{{Boyajian} {et~al.}(2009){Boyajian}, {McAlister}, {Cantrell}, {Gies},
  {Brummelaar}, {Farrington}, {Goldfinger}, {Sturmann}, {Sturmann}, {Turner},
  \& {Ridgway}}]{Boyajian2009}
{Boyajian}, T.~S., {McAlister}, H.~A., {Cantrell}, J.~R., {et~al.} 2009, \apj,
  691, 1243

\bibitem[{{Breger} {et~al.}(2006){Breger}, {Beck}, {Lenz}, {Schmitzberger},
  {Guggenberger}, \& {Shobbrook}}]{Breger2006}
{Breger}, M., {Beck}, P., {Lenz}, P., {et~al.} 2006, \aap, 455, 673

\bibitem[{{Butler} {et~al.}(2004){Butler}, {Bedding}, {Kjeldsen}, {McCarthy},
  {O'Toole}, {Tinney}, {Marcy}, \& {Wright}}]{Butler2004}
{Butler}, R.~P., {Bedding}, T.~R., {Kjeldsen}, H., {et~al.} 2004, \apjl, 600,
  L75

\bibitem[{{Carney} {et~al.}(2003){Carney}, {Latham}, {Stefanik}, {Laird}, \&
  {Morse}}]{Carney2003}
{Carney}, B.~W., {Latham}, D.~W., {Stefanik}, R.~P., {Laird}, J.~B., \&
  {Morse}, J.~A. 2003, \aj, 125, 293

\bibitem[{{Carrier} {et~al.}(2003){Carrier}, {Bouchy}, \&
  {Eggenberger}}]{Carrier2003}
{Carrier}, F., {Bouchy}, F., \& {Eggenberger}, P. 2003, in Asteroseismology
  Across the HR Diagram, ed. M.~J. {Thompson}, M.~S. {Cunha}, \& M.~J.~P.~F.~G.
  {Monteiro}, 311--314

\bibitem[{{Casagrande} {et~al.}(2014){Casagrande}, {Portinari}, {Glass},
  {Laney}, {Silva Aguirre}, {Datson}, {Andersen}, {Nordstr{\"o}m}, {Holmberg},
  {Flynn}, \& {Asplund}}]{Casagrande2014}
{Casagrande}, L., {Portinari}, L., {Glass}, I.~S., {et~al.} 2014, \mnras, 439,
  2060

\bibitem[{{Chaplin} {et~al.}(2011){Chaplin}, {Kjeldsen},
  {Christensen-Dalsgaard}, {Basu}, {Miglio}, {Appourchaux}, {Bedding},
  {Elsworth}, {Garc{\'{\i}}a}, {Gilliland}, {Girardi}, {Houdek}, {Karoff},
  {Kawaler}, {Metcalfe}, {Molenda-{\.Z}akowicz}, {Monteiro}, {Thompson},
  {Verner}, {Ballot}, {Bonanno}, {Brand{\~a}o}, {Broomhall}, {Bruntt},
  {Campante}, {Corsaro}, {Creevey}, {Do{\u g}an}, {Esch}, {Gai}, {Gaulme},
  {Hale}, {Handberg}, {Hekker}, {Huber}, {Jim{\'e}nez}, {Mathur}, {Mazumdar},
  {Mosser}, {New}, {Pinsonneault}, {Pricopi}, {Quirion}, {R{\'e}gulo},
  {Salabert}, {Serenelli}, {Silva Aguirre}, {Sousa}, {Stello}, {Stevens},
  {Suran}, {Uytterhoeven}, {White}, {Borucki}, {Brown}, {Jenkins}, {Kinemuchi},
  {Van Cleve}, \& {Klaus}}]{Chaplin2011}
{Chaplin}, W.~J., {Kjeldsen}, H., {Christensen-Dalsgaard}, J., {et~al.} 2011,
  Science, 332, 213

\bibitem[{{Chelli} {et~al.}(2009){Chelli}, {Utrera}, \&
  {Duvert}}]{2009AAChelli}
{Chelli}, A., {Utrera}, O.~H., \& {Duvert}, G. 2009, \aap, 502, 705

\bibitem[{{Choi} {et~al.}(1995){Choi}, {Soon}, {Donahue}, {Baliunas}, \&
  {Henry}}]{Choi1995}
{Choi}, H.-J., {Soon}, W., {Donahue}, R.~A., {Baliunas}, S.~L., \& {Henry},
  G.~W. 1995, \pasp, 107, 744

\bibitem[{{Claret} \& {Bloemen}(2011)}]{ClaretBloemen2011}
{Claret}, A. \& {Bloemen}, S. 2011, A\&A, 529, A75

\bibitem[{{Corsaro} {et~al.}(2012{\natexlab{a}}){Corsaro}, {Grundahl},
  {Leccia}, {Bonanno}, {Kjeldsen}, \& {Patern{\`o}}}]{corsaro2012b}
{Corsaro}, E., {Grundahl}, F., {Leccia}, S., {et~al.} 2012{\natexlab{a}}, \aap,
  537, A9

\bibitem[{{Corsaro} {et~al.}(2012{\natexlab{b}}){Corsaro}, {Stello}, {Huber},
  {Bedding}, {Bonanno}, {Brogaard}, {Kallinger}, {Benomar}, {White}, {Mosser},
  {Basu}, {Chaplin}, {Christensen-Dalsgaard}, {Elsworth}, {Garc{\'{\i}}a},
  {Hekker}, {Kjeldsen}, {Mathur}, {Meibom}, {Hall}, {Ibrahim}, \&
  {Klaus}}]{corsaro2012}
{Corsaro}, E., {Stello}, D., {Huber}, D., {et~al.} 2012{\natexlab{b}}, ApJ,
  757, 190

\bibitem[{{De Ridder} {et~al.}(2009){De Ridder}, {Barban}, {Baudin}, {Carrier},
  {Hatzes}, {Hekker}, {Kallinger}, {Weiss}, {Baglin}, {Auvergne}, {Samadi},
  {Barge}, \& {Deleuil}}]{deRidder2009}
{De Ridder}, J., {Barban}, C., {Baudin}, F., {et~al.} 2009, Nature, 459, 398

\bibitem[{{Degroote} {et~al.}(2011){Degroote}, {Acke}, {Samadi}, {Aerts},
  {Kurtz}, {Noels}, {Miglio}, {Montalb{\'a}n}, {Bloemen}, {Baglin}, {Baudin},
  {Catala}, {Michel}, \& {Auvergne}}]{2011AADegroote}
{Degroote}, P., {Acke}, B., {Samadi}, R., {et~al.} 2011, \aap, 536, A82

\bibitem[{{Deheuvels} {et~al.}(2012){Deheuvels}, {Garc{\'{\i}}a}, {Chaplin},
  {Basu}, {Antia}, {Appourchaux}, {Benomar}, {Davies}, {Elsworth}, {Gizon},
  {Goupil}, {Reese}, {Regulo}, {Schou}, {Stahn}, {Casagrande},
  {Christensen-Dalsgaard}, {Fischer}, {Hekker}, {Kjeldsen}, {Mathur}, {Mosser},
  {Pinsonneault}, {Valenti}, {Christiansen}, {Kinemuchi}, \&
  {Mullally}}]{deheuvels2012}
{Deheuvels}, S., {Garc{\'{\i}}a}, R.~A., {Chaplin}, W.~J., {et~al.} 2012, ApJ,
  756, 19

\bibitem[{{Desmet} {et~al.}(2009){Desmet}, {Briquet}, {Thoul}, {Zima}, {De
  Cat}, {Handler}, {Ilyin}, {Kambe}, {Krzesinski}, {Lehmann}, {Masuda},
  {Mathias}, {Mkrtichian}, {Telting}, {Uytterhoeven}, {Yang}, \&
  {Aerts}}]{Desmet2009}
{Desmet}, M., {Briquet}, M., {Thoul}, A., {et~al.} 2009, \mnras, 396, 1460

\bibitem[{{Deubner}(1975)}]{Deubner1975}
{Deubner}, F.-L. 1975, \aap, 44, 371

\bibitem[{{Elsworth} {et~al.}(1995){Elsworth}, {Howe}, {Isaak}, {McLeod},
  {Miller}, {New}, {Wheeler}, \& {Gough}}]{elsworth1995}
{Elsworth}, Y., {Howe}, R., {Isaak}, G.~R., {et~al.} 1995, Nature, 376, 669

\bibitem[{{Flower}(1996)}]{Flower1996}
{Flower}, P.~J. 1996, \apj, 469, 355

\bibitem[{{Frandsen} {et~al.}(2002){Frandsen}, {Carrier}, {Aerts}, {Stello},
  {Maas}, {Burnet}, {Bruntt}, {Teixeira}, {de Medeiros}, {Bouchy}, {Kjeldsen},
  {Pijpers}, \& {Christensen-Dalsgaard}}]{frandsen2002}
{Frandsen}, S., {Carrier}, F., {Aerts}, C., {et~al.} 2002, A\&A, 394, L5

\bibitem[{{Frandsen} {et~al.}(2013){Frandsen}, {Lehmann}, {Hekker},
  {Southworth}, {Debosscher}, {Beck}, {Hartmann}, {Pigulski}, {Kopacki},
  {Ko{\l}aczkowski}, {St{\c e}{\'s}licki}, {Thygesen}, {Brogaard}, \&
  {Elsworth}}]{frandsen2013}
{Frandsen}, S., {Lehmann}, H., {Hekker}, S., {et~al.} 2013, \aap, 556, A138

\bibitem[{{Garcia} {et~al.}(2014){Garcia}, {Ceillier}, {Salabert}, {Mathur},
  {van Saders}, {Pinsonneault}, {Ballot}, {Beck}, {Bloemen}, {Campante},
  {Davies}, {do Nascimento}, {Mathis}, {Metcalfe}, {Nielsen}, {Suarez},
  {Chaplin}, {Jimenez}, \& {Karoff}}]{Garcia2014a}
{Garcia}, R.~A., {Ceillier}, T., {Salabert}, D., {et~al.} 2014, ArXiv:
  1403.7155

\bibitem[{{Garc{\i}a} {et~al.}(2014){Garc{\i}a}, {Mathur}, {Pires}, {Regulo},
  {Bellamy}, {Palle}, {Ballot}, {Barcelo Forteza}, {Beck}, {Bedding},
  {Ceillier}, {Roca Cortes}, {Salabert}, \& {Stello}}]{Garcia2014b}
{Garc{\i}a}, R.~A., {Mathur}, S., {Pires}, S., {et~al.} 2014, ArXiv: 1405.5374

\bibitem[{{Gratton} {et~al.}(2001){Gratton}, {Bonanno}, {Bruno}, {Cali},
  {Claudi}, {Cosentino}, {Desidera}, {Diego}, {Farisato}, {Martorana},
  {Rebeschini}, \& {Scuderi}}]{SARG2001}
{Gratton}, R.~G., {Bonanno}, G., {Bruno}, P., {et~al.} 2001, Experimental
  Astronomy, 12, 107

\bibitem[{{Griffin}(2012)}]{Griffin2012}
{Griffin}, R.~F. 2012, Journal of Astrophysics and Astronomy, 33, 29

\bibitem[{{Grundahl} {et~al.}(2011){Grundahl}, {Christensen-Dalsgaard},
  {Gr{\aa}e J{\o}rgensen}, {Frandsen}, {Kjeldsen}, \& {Kj{\ae}rgaard
  Rasmussen}}]{Song2011}
{Grundahl}, F., {Christensen-Dalsgaard}, J., {Gr{\aa}e J{\o}rgensen}, U.,
  {et~al.} 2011, Journal of Physics Conference Series, 271, 012083

\bibitem[{{Hanbury Brown} {et~al.}(1974){Hanbury Brown}, {Davis}, {Lake}, \&
  {Thompson}}]{1974MNRASHanbury}
{Hanbury Brown}, R., {Davis}, J., {Lake}, R.~J.~W., \& {Thompson}, R.~J. 1974,
  \mnras, 167, 475

\bibitem[{{Handler} {et~al.}(2004){Handler}, {Shobbrook}, {Jerzykiewicz},
  {Krisciunas}, {Tshenye}, {Rodr{\'{\i}}guez}, {Costa}, {Zhou}, {Medupe},
  {Phorah}, {Garrido}, {Amado}, {Papar{\'o}}, {Zsuffa}, {Ramokgali}, {Crowe},
  {Purves}, {Avila}, {Knight}, {Brassfield}, {Kilmartin}, \&
  {Cottrell}}]{Handler2004}
{Handler}, G., {Shobbrook}, R.~R., {Jerzykiewicz}, M., {et~al.} 2004, \mnras,
  347, 454

\bibitem[{{Hekker} \& {Aerts}(2010)}]{hekkerAerts2010}
{Hekker}, S. \& {Aerts}, C. 2010, \aap, 515, A43

\bibitem[{{Hekker} {et~al.}(2006){Hekker}, {Aerts}, {De Ridder}, \&
  {Carrier}}]{hekker2006}
{Hekker}, S., {Aerts}, C., {De Ridder}, J., \& {Carrier}, F. 2006, A\&A, 458,
  931

\bibitem[{{Hekker} {et~al.}(2009){Hekker}, {Kallinger}, {Baudin}, {De Ridder},
  {Barban}, {Carrier}, {Hatzes}, {Weiss}, \& {Baglin}}]{Hekker2009}
{Hekker}, S., {Kallinger}, T., {Baudin}, F., {et~al.} 2009, \aap, 506, 465

\bibitem[{{Hekker} \& {Mel{\'e}ndez}(2007)}]{hekker2007}
{Hekker}, S. \& {Mel{\'e}ndez}, J. 2007, A\&A, 475, 1003

\bibitem[{{Huber} {et~al.}(2011){Huber}, {Bedding}, {Stello}, {Hekker},
  {Mathur}, {Mosser}, {Verner}, {Bonanno}, {Buzasi}, {Campante}, {Elsworth},
  {Hale}, {Kallinger}, {Silva Aguirre}, {Chaplin}, {De Ridder},
  {Garc{\'{\i}}a}, {Appourchaux}, {Frandsen}, {Houdek}, {Molenda-{\.Z}akowicz},
  {Monteiro}, {Christensen-Dalsgaard}, {Gilliland}, {Kawaler}, {Kjeldsen},
  {Broomhall}, {Corsaro}, {Salabert}, {Sanderfer}, {Seader}, \&
  {Smith}}]{huber2011}
{Huber}, D., {Bedding}, T.~R., {Stello}, D., {et~al.} 2011, ApJ, 743, 143

\bibitem[{{Huber} {et~al.}(2010){Huber}, {Bedding}, {Stello}, {Mosser},
  {Mathur}, {Kallinger}, {Hekker}, {Elsworth}, {Buzasi}, {De Ridder},
  {Gilliland}, {Kjeldsen}, {Chaplin}, {Garc{\'{\i}}a}, {Hale}, {Preston},
  {White}, {Borucki}, {Christensen-Dalsgaard}, {Clarke}, {Jenkins}, \&
  {Koch}}]{Huber2010}
{Huber}, D., {Bedding}, T.~R., {Stello}, D., {et~al.} 2010, \apj, 723, 1607

\bibitem[{{Izumiura}(1999)}]{Izumiura1999}
{Izumiura}, H. 1999, in Observational Astrophysics in Asia and its Future, ed.
  P.~S. {Chen}, 77

\bibitem[{{Kallinger} {et~al.}(2012){Kallinger}, {Hekker}, {Mosser}, {De
  Ridder}, {Bedding}, {Elsworth}, {Gruberbauer}, {Guenther}, {Stello}, {Basu},
  {Garc{\'{\i}}a}, {Chaplin}, {Mullally}, {Still}, \&
  {Thompson}}]{kallinger2012}
{Kallinger}, T., {Hekker}, S., {Mosser}, B., {et~al.} 2012, A\&A, 541, A51

\bibitem[{{Kallinger} {et~al.}(2010){Kallinger}, {Mosser}, {Hekker}, {Huber},
  {Stello}, {Mathur}, {Basu}, {Bedding}, {Chaplin}, {De Ridder}, {Elsworth},
  {Frandsen}, {Garc{\'{\i}}a}, {Gruberbauer}, {Matthews}, {Borucki}, {Bruntt},
  {Christensen-Dalsgaard}, {Gilliland}, {Kjeldsen}, \& {Koch}}]{kallinger2010}
{Kallinger}, T., {Mosser}, B., {Hekker}, S., {et~al.} 2010, A\&A, 522, A1

\bibitem[{{Kambe} {et~al.}(2008){Kambe}, {Ando}, {Sato}, {Izumiura}, {Sekii},
  {Paulson}, {Yanagisawa}, {Masuda}, {Shibahashi}, {Hatzes}, {Martic},
  {Lebrun}, {Mkrtichian}, {Kiss}, {Bruntt}, {O'Toole}, \&
  {Bedding}}]{Kambe2008}
{Kambe}, E., {Ando}, H., {Sato}, B., {et~al.} 2008, \pasj, 60, 45

\bibitem[{{Kambe} {et~al.}(2013){Kambe}, {Yoshida}, {Izumiura}, {Koyano},
  {Nagayama}, {Shimizu}, {Okada}, {Okita}, {Sakamoto}, {Sato}, \&
  {Yamamuro}}]{Kambe2013}
{Kambe}, E., {Yoshida}, M., {Izumiura}, H., {et~al.} 2013, \pasj, 65, 15

\bibitem[{{Kjeldsen} \& {Bedding}(1995)}]{kjeldsen1995}
{Kjeldsen}, H. \& {Bedding}, T.~R. 1995, A\&A, 293, 87

\bibitem[{{Kolenberg} {et~al.}(2009){Kolenberg}, {Guggenberger}, {Medupe},
  {Lenz}, {Schmitzberger}, {Shobbrook}, {Beck}, {Ngwato}, \&
  {Lub}}]{Kolenberg2009}
{Kolenberg}, K., {Guggenberger}, E., {Medupe}, T., {et~al.} 2009, \mnras, 396,
  263

\bibitem[{{Kurucz}(1993)}]{1993yCatKurucz}
{Kurucz}, R.~L. 1993, VizieR Online Data Catalog, 6039, 0

\bibitem[{{Le Bouquin} {et~al.}(2008){Le Bouquin}, {Abuter}, {Bauvir},
  {Bonnet}, {Haguenauer}, {di Lieto}, {Menardi}, {Morel}, {Rantakyr{\"o}},
  {Schoeller}, {Wallander}, \& {Wehner}}]{leBouquin2008}
{Le Bouquin}, J.-B., {Abuter}, R., {Bauvir}, B., {et~al.} 2008, in Society of
  Photo-Optical Instrumentation Engineers (SPIE) Conference Series, Vol. 7013,
  Society of Photo-Optical Instrumentation Engineers (SPIE) Conference Series

\bibitem[{{Lebreton} {et~al.}(2001){Lebreton}, {Fernandes}, \&
  {Lejeune}}]{Lebreton2001}
{Lebreton}, Y., {Fernandes}, J., \& {Lejeune}, T. 2001, \aap, 374, 540

\bibitem[{{Leighton} {et~al.}(1962){Leighton}, {Noyes}, \&
  {Simon}}]{Leighton1962}
{Leighton}, R.~B., {Noyes}, R.~W., \& {Simon}, G.~W. 1962, ApJ, 135, 474

\bibitem[{{Lenz} \& {Breger}(2005)}]{Lenz2005}
{Lenz}, P. \& {Breger}, M. 2005, Communications in Asteroseismology, 146, 53

\bibitem[{{Luck} \& {Heiter}(2007)}]{Luck2007}
{Luck}, R.~E. \& {Heiter}, U. 2007, \aj, 133, 2464

\bibitem[{{Lunt}(1919)}]{Lunt1919}
{Lunt}, J. 1919, \apj, 50, 161

\bibitem[{{McQuillan} {et~al.}(2014){McQuillan}, {Mazeh}, \&
  {Aigrain}}]{McQuillan2014}
{McQuillan}, A., {Mazeh}, T., \& {Aigrain}, S. 2014, \apjs, 211, 24

\bibitem[{{Moravveji} {et~al.}(2013){Moravveji}, {Guinan}, {Khosroshahi}, \&
  {Wasatonic}}]{Moravveji2012}
{Moravveji}, E., {Guinan}, E.~F., {Khosroshahi}, H., \& {Wasatonic}, R. 2013,
  \aj, 146, 148

\bibitem[{{Mosser} {et~al.}(2011){Mosser}, {Barban}, {Montalb{\'a}n}, {Beck},
  {Miglio}, {Belkacem}, {Goupil}, {Hekker}, {De Ridder}, {Dupret}, {Elsworth},
  {Noels}, {Baudin}, {Michel}, {Samadi}, {Auvergne}, {Baglin}, \&
  {Catala}}]{mosser2011a}
{Mosser}, B., {Barban}, C., {Montalb{\'a}n}, J., {et~al.} 2011, A\&A, 532, A86

\bibitem[{{Mosser} {et~al.}(2013){Mosser}, {Dziembowski}, {Belkacem}, {Goupil},
  {Michel}, {Samadi}, {Soszy{\'n}ski}, {Vrard}, {Elsworth}, {Hekker}, \&
  {Mathur}}]{Mosser2013b}
{Mosser}, B., {Dziembowski}, W.~A., {Belkacem}, K., {et~al.} 2013, \aap, 559,
  A137

\bibitem[{{Perryman} {et~al.}(1998){Perryman}, {Brown}, {Lebreton}, {Gomez},
  {Turon}, {Cayrel de Strobel}, {Mermilliod}, {Robichon}, {Kovalevsky}, \&
  {Crifo}}]{Perryman1998}
{Perryman}, M.~A.~C., {Brown}, A.~G.~A., {Lebreton}, Y., {et~al.} 1998, \aap,
  331, 81

\bibitem[{{Petrov} {et~al.}(2007){Petrov}, {Malbet}, {Weigelt}, {Antonelli},
  {Beckmann}, {Bresson}, {Chelli}, {Dugu{\'e}}, {Duvert}, {Gennari},
  {Gl{\"u}ck}, {Kern}, {Lagarde}, {Le Coarer}, {Lisi}, {Millour}, {Perraut},
  {Puget}, {Rantakyr{\"o}}, {Robbe-Dubois}, {Roussel}, {Salinari}, {Tatulli},
  {Zins}, {Accardo}, {Acke}, {Agabi}, {Altariba}, {Arezki}, {Aristidi},
  {Baffa}, {Behrend}, {Bl{\"o}cker}, {Bonhomme}, {Busoni}, {Cassaing},
  {Clausse}, {Colin}, {Connot}, {Delboulb{\'e}}, {Domiciano de Souza},
  {Driebe}, {Feautrier}, {Ferruzzi}, {Forveille}, {Fossat}, {Foy},
  {Fraix-Burnet}, {Gallardo}, {Giani}, {Gil}, {Glentzlin}, {Heiden},
  {Heininger}, {Hernandez Utrera}, {Hofmann}, {Kamm}, {Kiekebusch}, {Kraus},
  {Le Contel}, {Le Contel}, {Lesourd}, {Lopez}, {Lopez}, {Magnard}, {Marconi},
  {Mars}, {Martinot-Lagarde}, {Mathias}, {M{\`e}ge}, {Monin}, {Mouillet},
  {Mourard}, {Nussbaum}, {Ohnaka}, {Pacheco}, {Perrier}, {Rabbia}, {Rebattu},
  {Reynaud}, {Richichi}, {Robini}, {Sacchettini}, {Schertl}, {Sch{\"o}ller},
  {Solscheid}, {Spang}, {Stee}, {Stefanini}, {Tallon}, {Tallon-Bosc}, {Tasso},
  {Testi}, {Vakili}, {von der L{\"u}he}, {Valtier}, {Vannier}, \&
  {Ventura}}]{Petrov2007}
{Petrov}, R.~G., {Malbet}, F., {Weigelt}, G., {et~al.} 2007, A\&A, 464, 1

\bibitem[{{Queloz} {et~al.}(2000){Queloz}, {Mayor}, {Weber}, {Bl{\'e}cha},
  {Burnet}, {Confino}, {Naef}, {Pepe}, {Santos}, \& {Udry}}]{Queloz2000}
{Queloz}, D., {Mayor}, M., {Weber}, L., {et~al.} 2000, \aap, 354, 99

\bibitem[{{Raskin} {et~al.}(2011){Raskin}, {van Winckel}, {Hensberge},
  {Jorissen}, {Lehmann}, {Waelkens}, {Avila}, {de Cuyper}, {Degroote},
  {Dubosson}, {Dumortier}, {Fr{\'e}mat}, {Laux}, {Michaud}, {Morren}, {Perez
  Padilla}, {Pessemier}, {Prins}, {Smolders}, {van Eck}, \&
  {Winkler}}]{raskin2011}
{Raskin}, G., {van Winckel}, H., {Hensberge}, H., {et~al.} 2011, A\&A, 526, A69

\bibitem[{{Saesen} {et~al.}(2010){Saesen}, {Carrier}, {Pigulski}, {Aerts},
  {Handler}, {Narwid}, {Fu}, {Zhang}, {Jiang}, {Vanautgaerden}, {Kopacki},
  {St{\c e}{\'s}licki}, {Acke}, {Poretti}, {Uytterhoeven}, {Gielen},
  {{\O}stensen}, {De Meester}, {Reed}, {Ko{\l}aczkowski}, {Michalska},
  {Schmidt}, {Yakut}, {Leitner}, {Kalomeni}, {Cherix}, {Spano}, {Prins}, {van
  Helshoecht}, {Zima}, {Huygen}, {Vandenbussche}, {Lenz}, {Ladjal}, {Puga
  Antol{\'{\i}}n}, {Verhoelst}, {De Ridder}, {Niarchos}, {Liakos}, {Lorenz},
  {Dehaes}, {Reyniers}, {Davignon}, {Kim}, {Kim}, {Lee}, {Lee}, {Kwon},
  {Broeders}, {van Winckel}, {Vanhollebeke}, {Waelkens}, {Raskin}, {Blom},
  {Eggen}, {Degroote}, {Beck}, {Puschnig}, {Schmitzberger}, {Gelven},
  {Steininger}, {Blommaert}, {Drummond}, {Briquet}, \&
  {Debosscher}}]{Saesen2010}
{Saesen}, S., {Carrier}, F., {Pigulski}, A., {et~al.} 2010, \aap, 515, A16

\bibitem[{{Sato} {et~al.}(2007){Sato}, {Izumiura}, {Toyota}, {Kambe}, {Takeda},
  {Masuda}, {Omiya}, {Murata}, {Itoh}, {Ando}, {Yoshida}, {Ikoma}, {Kokubo}, \&
  {Ida}}]{Sato2007}
{Sato}, B., {Izumiura}, H., {Toyota}, E., {et~al.} 2007, \apj, 661, 527

\bibitem[{{Takeda} {et~al.}(2008){Takeda}, {Sato}, \& {Murata}}]{Takeda2008}
{Takeda}, Y., {Sato}, B., \& {Murata}, D. 2008, \pasj, 60, 781

\bibitem[{{Tallon-Bosc} {et~al.}(2008){Tallon-Bosc}, {Tallon}, {Thi{\'e}baut},
  {B{\'e}chet}, {Mella}, {Lafrasse}, {Chesneau}, {Domiciano de Souza},
  {Duvert}, {Mourard}, {Petrov}, \& {Vannier}}]{2008SPIETallonBosc}
{Tallon-Bosc}, I., {Tallon}, M., {Thi{\'e}baut}, E., {et~al.} 2008, in Society
  of Photo-Optical Instrumentation Engineers (SPIE) Conference Series, Vol.
  7013, Society of Photo-Optical Instrumentation Engineers (SPIE) Conference
  Series

\bibitem[{{Tatulli} {et~al.}(2007){Tatulli}, {Millour}, {Chelli}, {Duvert},
  {Acke}, {Hernandez Utrera}, {Hofmann}, {Kraus}, {Malbet}, {M{\`e}ge},
  {Petrov}, {Vannier}, {Zins}, {Antonelli}, {Beckmann}, {Bresson}, {Dugu{\'e}},
  {Gennari}, {Gl{\"u}ck}, {Kern}, {Lagarde}, {Le Coarer}, {Lisi}, {Perraut},
  {Puget}, {Rantakyr{\"o}}, {Robbe-Dubois}, {Roussel}, {Weigelt}, {Accardo},
  {Agabi}, {Altariba}, {Arezki}, {Aristidi}, {Baffa}, {Behrend}, {Bl{\"o}cker},
  {Bonhomme}, {Busoni}, {Cassaing}, {Clausse}, {Colin}, {Connot},
  {Delboulb{\'e}}, {Domiciano de Souza}, {Driebe}, {Feautrier}, {Ferruzzi},
  {Forveille}, {Fossat}, {Foy}, {Fraix-Burnet}, {Gallardo}, {Giani}, {Gil},
  {Glentzlin}, {Heiden}, {Heininger}, {Kamm}, {Kiekebusch}, {Le Contel}, {Le
  Contel}, {Lesourd}, {Lopez}, {Lopez}, {Magnard}, {Marconi}, {Mars},
  {Martinot-Lagarde}, {Mathias}, {Monin}, {Mouillet}, {Mourard}, {Nussbaum},
  {Ohnaka}, {Pacheco}, {Perrier}, {Rabbia}, {Rebattu}, {Reynaud}, {Richichi},
  {Robini}, {Sacchettini}, {Schertl}, {Sch{\"o}ller}, {Solscheid}, {Spang},
  {Stee}, {Stefanini}, {Tallon}, {Tallon-Bosc}, {Tasso}, {Testi}, {Vakili},
  {von der L{\"u}he}, {Valtier}, \& {Ventura}}]{2007AATatulli}
{Tatulli}, E., {Millour}, F., {Chelli}, A., {et~al.} 2007, \aap, 464, 29

\bibitem[{{Torres} {et~al.}(1997){Torres}, {Stefanik}, \&
  {Latham}}]{Torres1997}
{Torres}, G., {Stefanik}, R.~P., \& {Latham}, D.~W. 1997, \apj, 485, 167

\bibitem[{{van Leeuwen}(2007)}]{vanLeeuwen2007}
{van Leeuwen}, F. 2007, \aap, 474, 653

\bibitem[{{V{\'e}rinaud} \& {Cassaing}(2001)}]{piston2001}
{V{\'e}rinaud}, C. \& {Cassaing}, F. 2001, \aap, 365, 314

\bibitem[{{Walker} {et~al.}(2003){Walker}, {Matthews}, {Kuschnig}, {Johnson},
  {Rucinski}, {Pazder}, {Burley}, {Walker}, {Skaret}, {Zee}, {Grocott},
  {Carroll}, {Sinclair}, {Sturgeon}, \& {Harron}}]{Walker2003}
{Walker}, G., {Matthews}, J., {Kuschnig}, R., {et~al.} 2003, PASP, 115, 1023

\bibitem[{{Winget} {et~al.}(1990){Winget}, {Nather}, {Clemens}, {Provencal},
  {Kleinman}, {Bradley}, {Wood}, {Claver}, {Robinson}, {Grauer}, {Hine},
  {Fontaine}, {Achilleos}, {Marar}, {Seetha}, {Ashoka}, {O'Donoghue}, {Warner},
  {Kurtz}, {Martinez}, {Vauclair}, {Chevreton}, {Kanaan}, {Kepler},
  {Augusteijn}, {van Paradijs}, {Hansen}, \& {Liebert}}]{Winget1990}
{Winget}, D.~E., {Nather}, R.~E., {Clemens}, J.~C., {et~al.} 1990, ApJ, 357,
  630

\bibitem[{{Wittkowski} {et~al.}(2004){Wittkowski}, {Aufdenberg}, \&
  {Kervella}}]{2004AAWittkowski}
{Wittkowski}, M., {Aufdenberg}, J.~P., \& {Kervella}, P. 2004, \aap, 413, 711

\end{thebibliography}

\end{document}